\documentclass[a4paper,fleqn,usenatbib]{mnras}
\usepackage{amssymb}
\usepackage{dcolumn}
\usepackage{bm}
\usepackage{lscape}
\usepackage{graphicx}
\usepackage{subfigure}
\usepackage{diagbox}
\usepackage{hyperref}
\usepackage[hyphenbreaks]{breakurl}
\usepackage{rotating}
\usepackage{tikz}
\usepackage{setspace}
\usepackage{array} 
\usepackage{adjustbox}
\usepackage{siunitx}
\usepackage{comment}

\graphicspath{{Pics/}}

\begin{document}

\title[DEVILS: Mass Growth of Morphological Types]{Deep Extragalactic VIsible Legacy Survey (DEVILS): Stellar Mass Growth by Morphological Type since $z = 1$}

\author[Hashemizadeh et al.]{Abdolhosein Hashemizadeh$^{1}$\thanks{E-mail: abdolhosein.hashemizadeh@research.uwa.edu.au}, Simon P. Driver$^{1}$, Luke J.~M.~Davies$^1$,
	\newauthor
	Aaron S. G. Robotham$^1$, Sabine Bellstedt$^1$, Rogier A. Windhorst$^2$, Malcolm Bremer$^3$,
	\newauthor
	Steven Phillipps$^3$, Matt Jarvis$^{4,5}$, Benne W. Holwerda$^6$, Claudia del P. Lagos$^1$, 
	\newauthor
	Soheil Koushan$^1$, Malgorzata Siudek$^{7,8}$, Natasha Maddox$^9$, Jessica E. Thorne$^1$,
	\newauthor
	Pascal Elahi$^1$
	\newauthor
	 \\
	\\
$^1$ICRAR, The University of Western Australia, 35 Stirling Highway, Crawley WA 6009, Australia\\
$^2$School of Earth and Space Exploration, Arizona State University, Tempe, AZ 85287-1404\\
$^3$Astrophysics Group, School of Physics, University of Bristol, Tyndall Avenue, Bristol BS8 1TL, UK\\
$^4$Department of Astrophysics, University of Oxford, The Denys Wilkinson Building, Keble Road, Oxford OX1 3RH, UK\\
$^5$Department of Physics, University of the Western Cape, Bellville 7535, South Africa \\
$^6$Department of Physics and Astronomy, University of Louisville, Natural Science Building 102, 40292 KY Louisville, USA \\
$^7$Institut de F\'{\i}sica d'Altes Energies (IFAE), The Barcelona Institute of Science and Technology, 08193 Bellaterra (Barcelona), Spain \\
$^8$National Centre for Nuclear Research, ul. Hoza 69, 00-681 Warsaw, Poland \\
$^9$Faculty of Physics, Ludwig-Maximilians-Universit\"at, Scheinerstr. 1, 81679 Munich, Germany \\
}

\date{Accepted XXX. Received YYY; in original form ZZZ}

\pubyear{2021}


\label{firstpage}
\pagerange{\pageref{firstpage}--\pageref{lastpage}}
\maketitle

\begin{abstract}
Using high-resolution Hubble Space Telescope imaging data, we perform a visual morphological classification of $\sim 36,000$ galaxies at $z < 1$ in the DEVILS/COSMOS region. As the main goal of this study, we derive the stellar mass function (SMF) and stellar mass density (SMD) sub-divided by morphological types. We find that visual morphological classification using optical imaging is increasingly difficult at $z > 1$ as the fraction of irregular galaxies and merger systems (when observed at rest-frame UV/blue wavelengths) dramatically increases.
We determine that roughly two-thirds of the total stellar mass of the Universe today was in place by $z \sim 1$. Double-component galaxies dominate the SMD at all epochs and increase in their contribution to the stellar mass budget to the present day. 
Elliptical galaxies are the second most dominant morphological type and increase their SMD by $\sim 2.5$ times, while by contrast, the pure-disk population significantly decreases by $\sim 85\%$. According to the evolution of both high- and low-mass ends of the SMF, we find that mergers and in-situ evolution in disks are both present at $z < 1$, and conclude that double-component galaxies are predominantly being built by the in-situ evolution in disks (apparent as the growth of the low-mass end with time), while mergers are likely responsible for the growth of ellipticals (apparent as the increase of intermediate/high-mass end).
\end{abstract}


\setlength{\extrarowheight}{0pt}

\section{Introduction}

The galaxy population in the local Universe is observed to be bimodal. This bimodality manifests in multiple properties such as colour, morphology, metallicity, light profile shape and environment (e.g. \citealt{Kauffmann03}; \citealt{Baldry04}; \citealt{Brinchmann04}). This bimodality is also found to extend to earlier epochs (see e.g., \citealt{Strateva01}; \citealt{Hogg02}; \citealt{Bell04}; \citealt{Driver06}; \citealt{Taylor09}; \citealt{Brammer09}; \citealt{Williams09}; \citealt{Brammer11}) across many measurable parameters, for example in colour (e.g. \citealt{Cirasuolo07}; \citealt{Cassata08}; \citealt{Taylor15}), morphological type (e.g. \citealt{Kelvin14}; \citealt{Whitaker15}; \citealt{Moffett16a}; \citealt{Krywult17}), size (e.g. \citealt{Lange15}), and specific star formation rate (e.g. \citealt{Whitaker14}; \citealt{Renzini15} and the references therein). The inference from these observations is that there are likely two evolutionary pathways regulated by mass and environment giving rise to this bimodality (\citealt{Driver06}; \citealt{Scarlata07b}; \citealt{DeLucia07}; \citealt{Peng10}; \citealt{Trayford16}). However, it is unclear as to whether studying the global properties of galaxies or their individual morphological components can better explain the origin of this bimodality \citep{Driver13} or whether more complex astrophysics is required.  

In this study, we explore the origin of this bimodality by using the global properties of galaxies to study the evolution of their stellar mass function (SMF). The SMF, a statistical tool for measuring and constraining the evolution of the galaxy population, is defined as the number density of galaxies per logarithmic mass interval (\citealt{Schechter76}). The SMF of galaxies in the local Universe is now very well studied and found to be described by a two-component \cite{Schechter76} function with a characteristic cut-off mass of between $10^{10.6}$-$10^{11}\mathrm{M}_\odot$ and a steepening to lower masses (for example see: \citealt{Baldry08}; \citealt{Peng10}; \citealt{Baldry12}; \citealt{Kelvin14}; \citealt{Weigel16}; \citealt{Moffett16a}; \citealt{Wright17} ). Several studies have also investigated the evolution of the SMF at higher redshifts (\citealt{Pozzetti10}; \citealt{Muzzin13}; \citealt{Whitaker14}; \citealt{Leja15}; \citealt{Mortlock15}; \citealt{Wright18}; \citealt{Kawinwanichakij20}). A fingerprint of this bimodality is also observed in the double-component Schechter function required to fit the local SMF (e.g. \citealt{Baldry12}; \citealt{Wright18}). At least two distinct galaxy populations corresponding to \textit{star-forming} and \textit{passive} systems are thought to be the origin of this bimodal shape, with \textit{star-forming} systems dominating the low-mass tail and \textit{passive} galaxies dominating the high-mass ``hump'' of the SMF (\citealt{Baldry12}; \citealt{Muzzin13}; \citealt{Wright18}). 

Many previous studies have separated galaxies into two main populations of star forming and passive (or a proxy thereof), and measured their individual stellar mass assemblies (e.g. \citealt{Pozzetti10}; \citealt{Tomczak14}; \citealt{Leja15}; \citealt{Davidzon17}). For example, by separating their sample into early- and late-type galaxies based on colour, \cite{Vergani08} studied the SMF of the samples and confirmed that $\sim 50\%$ of the red sequence galaxies were already formed by $z \sim 1$. 
\cite{Pannella06} used a sample of $\sim 1600$ galaxies split into early- intermediate- and late-type galaxies classified according to their S\'ersic index. Similar work was carried out by \cite{Bundy05} and \cite{Ilbert10}. In the local Universe, some studies have classified small samples of galaxies and investigated the SMF of different morphological types (\citealt{Fukugita07} and \citealt{Bernardi10} using The Sloan Digital Sky Survey, SDSS). For example \cite{Bernardi10} reported that elliptical galaxies contain 25\% of the total stellar mass density and 20\% of the luminosity density of the local Universe.

Splitting galaxies into only two broad populations of star-forming and passive (or early- and late-type) galaxies might be inadequate to comprehensively study all aspects of galaxy evolution (e.g., \citealt{Siudek18}). Most previous studies at higher redshifts have separated galaxies into early- and late-type according to their colour distribution mainly due to the lack of spatial resolution as at high redshifts galaxies become unresolved in ground-based imaging. This has restricted true morphological comparisons in most studies to the local Universe (see e.g., \citealt{Fontana04}; \citealt{Baldry04}). A disadvantage of a simple colour-based classification is at higher redshifts, where for example regular disks occupied by old stellar populations are likely to be classified as early-type systems (see e.g., \citealt{Pannella06}). This is however unlikely to have an impact on the visual morphological classification, particularly when using high resolution imaging such as the Hubble Space Telescope (HST) data (\citealt{Huertas-Company15}). As such, we need better ways of quantifying the structure of galaxies throughout the history of the Universe. 
Consequently, some studies have started to probe visually classified morphological types (e.g., \citealt{Sandage05}; \citealt{Fukugita07}; \citealt{Nair10}; \citealt{Lintott11} and the references therein). In effect, there are two morphologies one might track: the end-point morphology at $z \sim 0$ or the instantaneous morphology at the observed epoch. Observing galaxies at higher redshifts we find a different distribution of morphological types. For example, in earlier works using HST, \cite{vandenBergh96} and \cite{Abraham96} presented a morphological catalogue of galaxies in the Hubble Deep Field and found significantly more interacting, merger and asymmetric galaxies than in the nearby Universe.

By studying the SMF of various morphological types in the local Universe ($z < 0.06$), visually classified in the Galaxy and Mass Assembly survey (GAMA, \citealt{Driver11}), \cite{Kelvin14} and \cite{Moffett16a} found that the local stellar mass density is dominated by spheroidal galaxies, defined as E/S0/Sa. They reported the contribution of spheroid- (E/S0ab) and disk-dominated (Sbcd/Irr) galaxies of approximately 70\% and 30\%, respectively, towards the total stellar mass budget of the local Universe. These studies, however, are limited to the very local Universe. For a better understanding of the galaxy formation and evolution processes at play in the evolution of different morphological types, and their rates of action, we need to extend similar analyses to higher redshifts. This will allow us to explore the contribution of different morphological types to the stellar mass budget as a function of time and elucidate the galaxy evolution process.

In the present work, we perform a visual morphological classification of galaxies in the DEVILS-D10/COSMOS field (\citealt{Scoville07}; \citealt{Davies18}). We make use of the high resolution HST imaging and use a sample of galaxies within $0 \le z \le 1$ and $\mathrm{M}_* > 10^{9.5}$, selected from the DEVILS sample and analysis, for which classification is reliable. This intermediate redshift range is a key phase in the evolution of the Universe where a large fraction ($\sim 50\%$) of the present-day stellar mass is formed and large structures such as groups, clusters and filaments undergo a significant evolution (\citealt{Davies18}). During this period the sizes and masses of galaxies also appear to undergo significant evolution (e.g. \citealt{vanderWel12}). 

By investigating the evolution of the SMF for different morphological types across cosmic time we can probe the various systems within which stars are located, and how they evolve. In this paper, we investigate the evolution of the stellar mass function of different morphological types and explore the contribution of each morphology to the global stellar mass build-up of the Universe. Shedding light on the redistribution of stellar mass in the Universe and also the transformation and redistribution of the stellar mass between different morphologies likely explains the origin of the bimodality in galaxy populations observed in the local Universe. 

The ultimate goal of this study and its companion papers is to not only probe the evolution of the morphological types but also study the formation and evolution of the galaxy structures, including bulges and disks. This will be presented in a companion paper (Hashemizadeh et al. in prep.), while in this paper we explore the visual morphological evolution of galaxies and conduct an assessment into the possibility of the bulge formation scenarios by constructing the global stellar mass distributions and densities of various morphological types. 

The data that we use in this work are presented in Section \ref{sec:AvailData}. In Section \ref{subsec:SampleSel}, we define our sample and sample selection. Section \ref{sec:MorphClass} presents different methods that we explore for the morphological classification. In Section \ref{sec:FitSMF}, we describe the parameterisation of the SMF. We then show the \textit{total} SMF at low-$z$ in Section \ref{subsec:MfunZ0}. Section \ref{subsec:LSS_cor} describes the effects of the cosmic large scale structure due to the limiting size of the DEVILS/COSMOS field, and the technique we use to correct for this. The evolution of the SMF and the SMD and their subdivision by morphological type are discussed in Sections \ref{subsec:MfunEvol}-\ref{sec:rho}. We finally discuss and summarize our results in Sections \ref{sec:discussion}-\ref{sec:summary}.

Throughout this paper, we use a flat standard $\Lambda$CDM cosmology of $\Omega_{\mathrm{M}} = 0.3$, $\Omega_\Lambda = 0.7$ with $H_0 = 70 \mathrm{km}\mathrm{s}^{-1}\mathrm{Mpc}^{-1}$. Magnitudes are given in the AB system.

\subsection{DEVILS: Deep Extragalactic VIsible Legacy Survey}
\label{sec:DEVILS}

The Deep Extragalactic VIsible Legacy Survey (DEVILS) \citep{Davies18}, is an ongoing magnitude-limited spectroscopic and multi-wavelength survey. Spectroscopic observations are currently being undertaken at the Anglo-Australian Telescope (AAT), providing spectroscopic redshift completeness of $> 95\%$ to Y-mag $< 21.2$ mag. 
This spectroscopic sample is supplemented with robust photometric redshifts, newly derived photometric catalogues (Davies et al. in prep.) and derived physical properties (\citealt{Thorne20}, and the work here) all undertaken as part of the DEVILS project. These catalogues extend to Y$\sim25$ mag in the D10 region.

The objective of the DEVILS campaign is to obtain a sample with high spectroscopic completeness extending over intermediate redshifts ($0.3 < z < 1.0$). At present, there is a lack of high completeness spectroscopic data in this redshift range (e.g., \citealt{Davies18}; \citealt{Scodeggio18}). DEVILS will fill this gap and allow for the construction of group, pair and filamentary catalogues to fold in environmental metrics. The DEVILS campaign covers three well-known fields: the XMM-Newton Large-Scale Structure field (D02: XMM-LSS), the Extended Chandra Deep Field-South (D03: ECDFS), and the Cosmological Evolution Survey field (D10: COSMOS). In this work, we only explore the D10 region as this overlaps with the HST COSMOS imaging. The spectroscopic redshifts used in this paper are from the DEVILS combined spectroscopic catalogue which includes all available redshifts in the COSMOS region, including zCOSMOS (\citealt{Lilly07}, \citealt{Davies15a}), hCOSMOS \citep{Damjanov18} and DEVILS \citep{Davies18}. As the DEVILS survey is ongoing the spectroscopic observations for our full sample are still incomplete (the completeness of the DEVILS combined data is currently $\sim 90$ per cent to Y-mag $= 20$). Those objects without spectroscopic redshifts are assigned photometric redshifts in the DEVILS master redshift catalogue (\texttt{DEVILS\_D10MasterRedshiftCat\_v0.2} catalogue), described in detail in \cite{Thorne20}. In this work, we also use the stellar mass measurements for the D10 region (\texttt{DEVILS\_D10ProSpectCat\_v0.3} catalogue) reported by \cite{Thorne20}.   
Briefly, to estimate stellar masses they used the {\sc ProSpect} SED fitting code \citep{Robotham20} adopting \cite{BC03} stellar libraries, the \cite{Chabrier03} IMF together with \cite{Charlot00} to model dust attenuation and \cite{Dale14} to model dust emission. This study makes use of the new multiwavelength photometry catalogue in the D10 field (\texttt{DEVILS\_PhotomCat\_v0.4}; Davies et al. in prep.) and finds stellar masses $\sim 0.2$ dex higher than in COSMOS2015 \citep{Laigle16} due to differences in modelling an evolving gas phase metallicity. See \cite{Thorne20} for more details.

\begin{figure}
	\centering
	\includegraphics[width = 0.53 \textwidth]{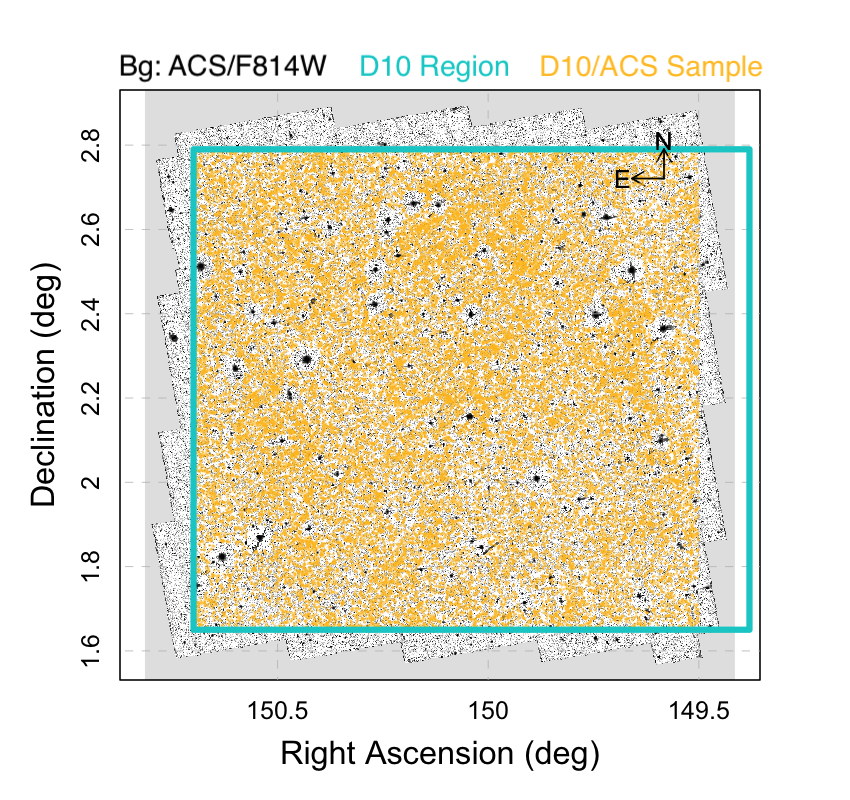}
	\caption{ Background shows the ACS/F814W mosaic image of the COSMOS field. The cyan rectangle represents the D10 region in the DEVILS survey. Gold points are the D10/ACS sources used in this work consisting of $\sim 36$k galaxies. See section \ref{subsec:SampleSel} for more detail. 
    }
	\label{fig:F814W_DEVILS_smp_z1}
\end{figure}

\subsection{COSMOS ACS/WFC imaging data}
\label{subsec:ACS/HST}

The Cosmic Evolution Survey (COSMOS) is one of the most comprehensive deep-field surveys to date, covering almost 2 contiguous square degrees of sky, designed to explore large scale structures and the evolution of galaxies, AGN and dark matter \citep{Scoville07}. The high resolution Hubble Space Telescope (HST) F814W imaging in COSMOS allows for the study of galaxy morphology and structure out to the detection limits. In total COSMOS detects $\sim 2$ million galaxies at a resolution of $< 100$ pc \citep{Scoville07}. 

The COSMOS region is centred at RA = $150.121$ ($10:00:28.600$) and DEC = $+2.21$ ($+02:12:21.00$) (J2000) and is supplemented by 1.7 square degrees of imaging with the Advanced Camera for Surveys (ACS\footnote{ACS Hand Book: \href{http://www.stsci.edu/hst/acs/documents/handbooks/current/c05\_imaging7.html\#357803}{www.stsci.edu/hst/acs/documents/}}) on HST. This 1.7 square degree region was observed during 590 orbits in the F814W (I-band) filter and also, 0.03 square degrees with F475W (g-band). In this study we exclusively use the F814W filter, not only providing coverage but also suitable rest-frame wavelength for the study of optical morphology of galaxies out to $z\sim1$ \citep{Koekemoer07}. The original ACS pixel scale is 0.05 arcsec and consists of a series of overlapping pointings. These have been drizzled and re-sampled to 0.03 arcsec resolution using the MultiDrizzle code (\citealt{Koekemoer03}), which is the imaging data we use in this work. The frames were downloaded from the public NASA/IPAC Infrared Science Archive (IRSA) webpage\footnote{\href{https://irsa.ipac.caltech.edu/data/COSMOS/images/acs\_2.0/I/}{irsa.ipac.caltech.edu/data/COSMOS/images/acs\_2.0/I/} } as fits images.

\begin{figure} 
	\centering
	\includegraphics[width = 0.48\textwidth]{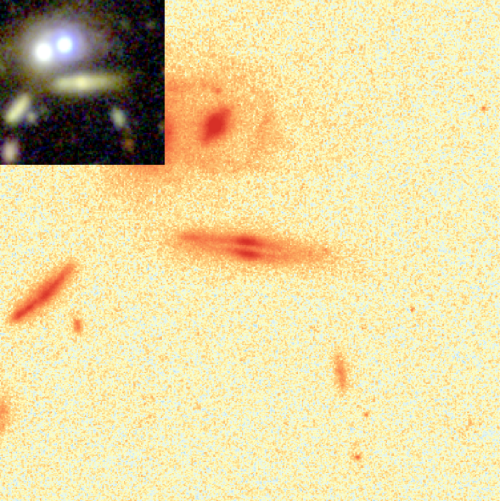}
	\caption{ A random galaxy in our sample at redshift $z \sim 0.47$ showing an example of the postage stamps we generate to perform our visual inspection. Main image is the HST ACS/F814W and the inset is SUBARU $gri$ colour image. The image shows a cutout of $5\times \mathrm{R90}$ on each side, where $\mathrm{R90} = 3.23$ arcsec measured from UltraVISTA Y-band \citep{Davies18}. }
	\label{fig:smplPostageStamp}
\end{figure}

\subsection{Sample selection: D10/ACS Sample} 
\label{subsec:SampleSel}

As part of the DEVILS survey we have selected an HST imaging sample with which to perform various morphological and structural science projects. 
Figure \ref{fig:F814W_DEVILS_smp_z1}, shows the ACS mosaic imaging of the COSMOS field with the full D10 region overlaid as a cyan rectangle (defined from the Ultra VISTA imaging). Our final sample, D10/ACS, is the common subset of sources from ACS and D10. The position of the D10/ACS sample on the plane of RA and DEC is overplotted on the same figure as yellow dots. Note that as shown in Figure \ref{fig:F814W_DEVILS_smp_z1}, we exclude objects in the jagged area of the ACS imaging leading to a rectangular effective area of $1.3467$ square degrees. 

Although our HST imaging data is high resolution (see Figure \ref{fig:smplPostageStamp} for a random galaxy at redshift $z \sim 0.47$), we are still unable to explore galaxy morphologies beyond a certain redshift and stellar mass in our analysis as galaxies become too small in angular scale or too dim in surface brightness to identify morphological substructures.

\begin{landscape}
\begin{figure} 
	\centering
	\includegraphics[width=1.3\textheight]{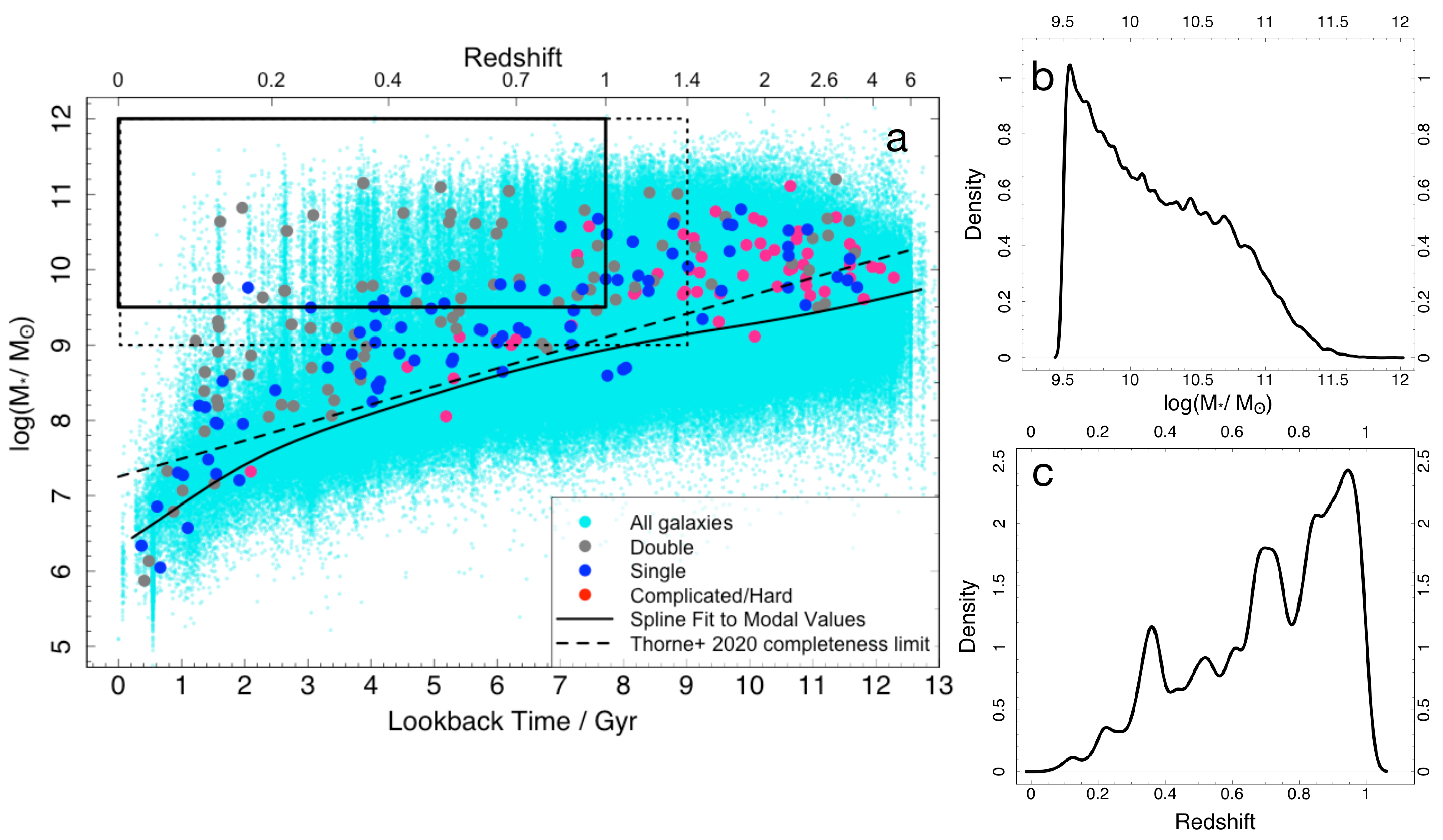}
	\caption{ (a) The relation between stellar mass and lookback time (redshift) of our sample. All galaxies are shown in cyan in background. Circles represent 284 galaxies that we randomly sample for initial visual inspection. Grey and blue circles show double and single component galaxies, respectively. Red symbols are complicated galaxies which consist of merging systems, perturbed galaxies, low S/N or high redshift clumpy galaxies. The dotted rectangle corresponds to our initial sample region ($z < 1.4$ and log$(\mathrm{M}_*/\mathrm{M}_{\odot}) > 9$). The solid rectangle shows our final sample region which covers galaxies up to $z < 1.0$ and log$(\mathrm{M}_*/\mathrm{M}_{\odot}) > 9.5$. The solid black line represents the spline fit to the modal value of the stellar mass in bins of lookback time indicating the stellar mass completeness. Black dashed line shows \protect\cite{Thorne20} completeness limit. See text for more details. Panels (b) and (c) display the distribution of stellar mass and redshift of our final sample (i.e. within the solid rectangle). Note that the PDFs are smoothed by a kernel with standard deviation of 0.02. }
	\label{fig:mzsmplsel}
\end{figure}
\end{landscape}

We first try to select a complete galaxy sample from the DEVILS-D10/COSMOS data to define the redshift and stellar mass range for which we can perform robust morphological classification and structural decomposition (Hashemizadeh et al. in prep.). Note that we make use of a combination of photometric and available spectroscopic redshifts as well as stellar masses as described in Section \ref{sec:DEVILS}. In total, at the time of writing this paper, $23,264$ spectroscopic redshifts are available in the D10 region (excluding the jagged edges) out of which $2,903$ redshifts are observed by DEVILS. See the DEVILS website\footnote{\href{https://devilsurvey.org}{https://devilsurvey.org}} for a full description of these data.
We select 284 random galaxies drawn from across the entire redshift and stellar mass distribution, and visually inspect them. These 284 galaxies are shown as circles in Figure \ref{fig:mzsmplsel}. From our visual inspection, we identify the boundaries within which we believe the majority of galaxies are sufficiently resolved that morphological classifications should be possible (i.e., not too small or faint). Our visual assessments are indicated by colour in Figure \ref{fig:mzsmplsel} showing two-component (grey), single component (blue), and problematic cases (red; merger, disrupted, and low S/N). Unsurprisingly, in agreement with other studies (e.g. \citealt{Conselice05}), we find the fraction of problematic galaxies increases drastically at high redshifts ($z > 1.4$). Beyond this redshift an increasing number of galaxies ($\sim 50\%$) become complex and no longer adhere to a simplistic picture of a central bulge plus disk system. A large fraction of galaxies appear interacting, clumpy, very faint and/or extremely \textit{compact}, hence the notion of galaxies as predominantly bulge plus disk systems becomes untenable. Note that the vast majority of galaxies in this redshift range are disturbed interacting systems, however our observation of a fraction of the clumpy galaxies could be due to the fact that the bluer rest-frame emission is more sensitive to star forming regions dominating the flux. The clumps and potential bulges at this epoch, are mostly comparable or smaller than the HST PSF, hence conventional 2D bulge+disk fits are also unlikely to be credible even at HST resolution. 

\begin{figure*}
	\centering
	\includegraphics[width = \textwidth]{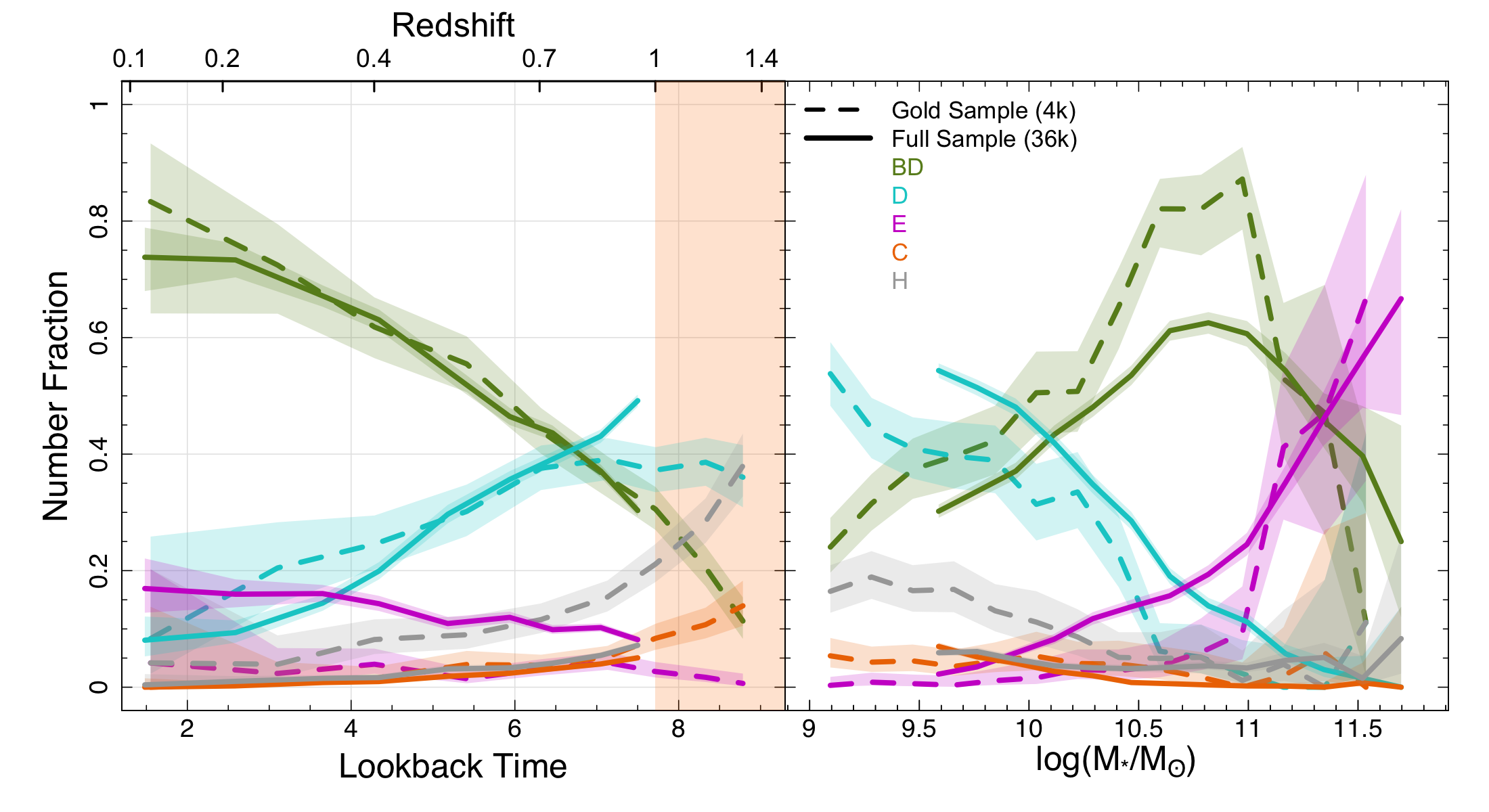}
	\caption{ The number fraction of each morphological type (BD: \textit{bulge+disk}, D: \textit{pure-disk}, E: \textit{elliptical}, C: \textit{compact}, H: \textit{hard}) in bins of lookback time (redshift) and total stellar mass, left and right panels, respectively. Each colour represents a morphology as indicated in the inset legend. Dashed lines are the $\sim4000$ random sample that we visually classified ($z < 1.4$) while solid lines are our full visual inspection of the final sample. Shaded stripes around lines display 95\% confidence intervals from beta distribution method as calculated using {\fontfamily{qcr}\selectfont prop.test} in R package {\fontfamily{qcr}\selectfont stats}.}
	\label{fig:frac_z1}
\end{figure*}

\begin{figure*}
	\centering
	\includegraphics[width=\textwidth, height=11.5cm]{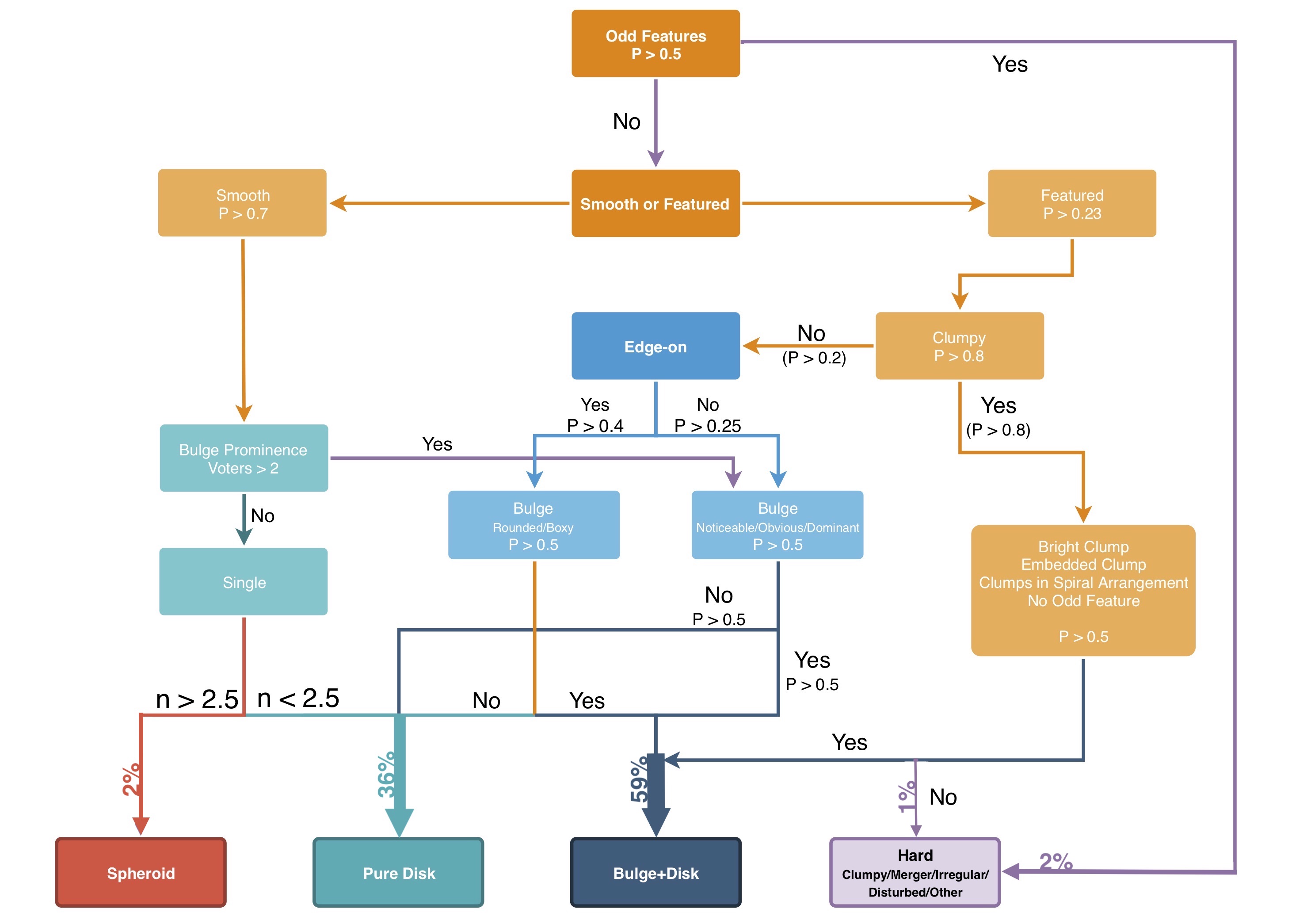}
	\caption{ This flowchart shows the decision tree that we adopt to translate Galaxy Zoo: Hubble (GZH) tasks and outputs into the desired morphologies to be used in this study. The weight of final arrows are proportional to the number of galaxies following those paths. In addition, the fraction of galaxies falling into each morphological category is annotated.}
	\label{fig:GZH}
\end{figure*}

\begin{figure*}
	\centering
	\includegraphics[width=\textwidth, height=11.5cm]{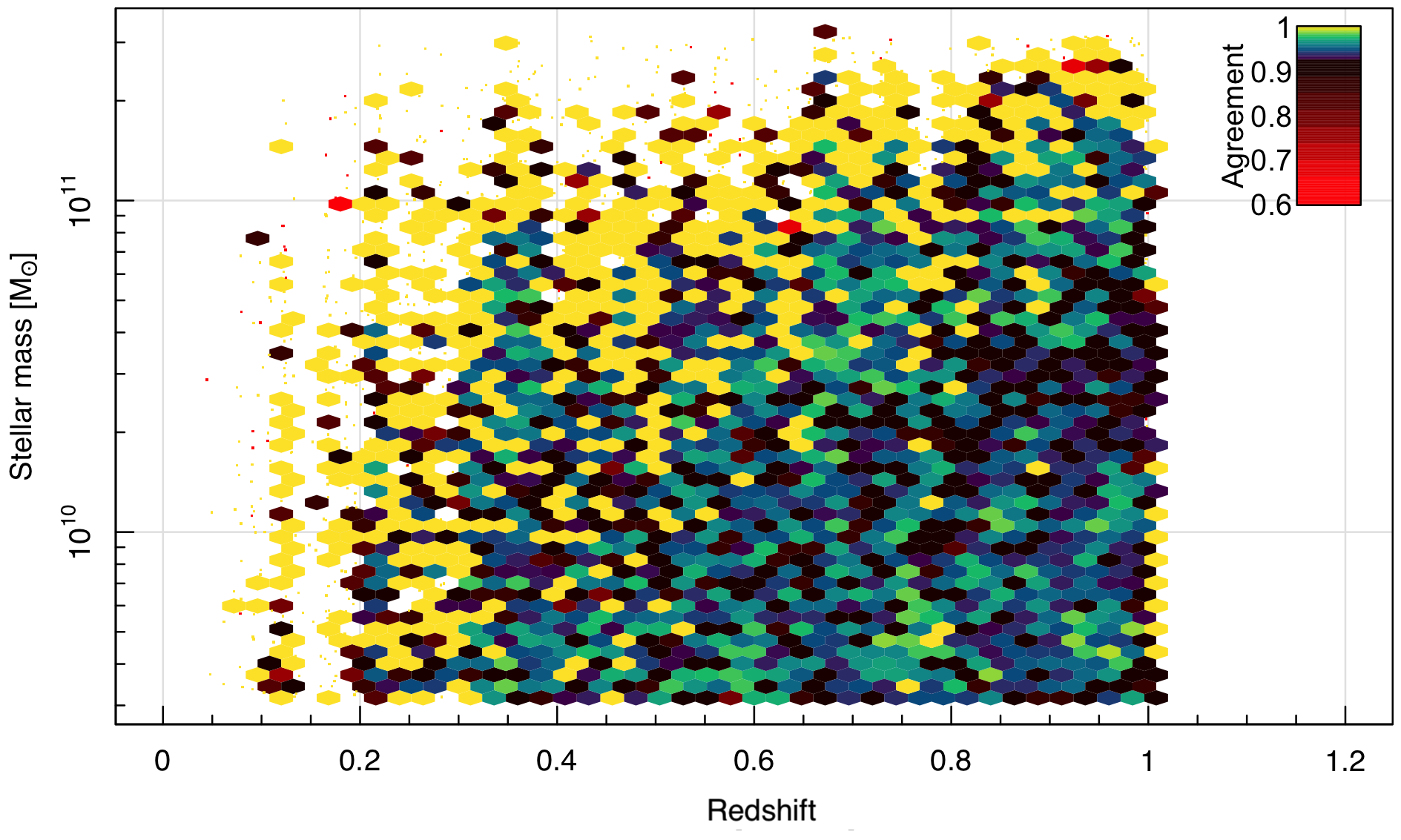}
	\caption{Stellar mass versus redshift, colour coded by the level of agreement in our final classifications (by three co-authors), i.e. the percentage of objects in each cell consistently classified by at least two of our classifiers. Note that we increase the resolution of the colour map of agreement $\gtrsim 0.92$ to highlight the variation of agreement in this level. See the text for more details.}
	\label{fig:agreent}
\end{figure*}

Note that as can be seen in Figure \ref{fig:mzsmplsel}, we also start to suffer significant mass incompleteness at low redshifts due to the COSMOS F814W sensitivity limit. The sample completeness limit is shown here as a smooth spline fitted to the modal values of stellar mass in bins of lookback time (shown as a solid line), i.e., peaks of the stellar mass histograms in narrow redshift slices. We also show the completeness limit reported by \cite{Thorne20} and shown as a dashed line in Figure \ref{fig:mzsmplsel}, which is based on a $g-i$ colour analysis and is formulated as $\mathrm{log(M_*)} = 0.25t + 7.25$, where $t$ is lookback time in Gyr. From this initial inspection we define our provisional window of 105,185 galaxies at $z < 1.4$ and log$(\mathrm{M}_*/\mathrm{M}_{\odot}) > 9$, shown as a dotted rectangle in Figure \ref{fig:mzsmplsel}.

To further tune this selection, we then generated postage stamps of $4,000$ random galaxies (now initial selection) using the HST F814W imaging and colour $gri$ insets from the Subaru Suprime-Cam data \citep{Taniguchi07}. Figure \ref{fig:smplPostageStamp} shows an example of the cutouts we generated for our visual inspection. The postage stamps are generated with $5 \times \mathrm{R90}$ on each side, where $\mathrm{R90}$ is the radius enclosing 90\% of the total flux in the UltraVISTA Y-band (soon to be presented in DEVILS photometry catalogue, Davies et al. in prep.).  
These stamps were independently reviewed by five authors (AH, SPD, ASGR, LJMD, SB) and classified into single component, double component and complicated systems (hereafter: \textit{hard}). The single component systems were later subdivided into disk or elliptical systems. Note that the \textit{hard} class consists of asymmetric, merging, clumpy, extremely compact and low-S/N systems, for which 2D structural decomposition would be unlikely to yield meaningful output. Objects with three or more votes in one category were adopted and more disparate outcomes discussed and debated until an agreement was obtained. In this way, we established a ``gold calibration sample" of 4k galaxies to justify our final redshift and stellar mass range, and for later use as a training sample in our automated-classification process, see Section \ref{sec:MorphClass} for more details. 

Figure \ref{fig:frac_z1} shows the fraction of each of the above classifications versus redshift and total stellar mass (dashed lines). As the left panel shows, the fraction of \textit{hard} galaxies (gray dashed line) drastically increases at $z > 1$. At the highest redshift of our sample, $z \sim 1.4$, 40 percent of the galaxies are deemed unfittable, or at least inconsistent with the notion of a classical/pseudo-bulge plus disk systems (this is consistent with \citealt{Abraham96b} and \citealt{Conselice05}). Also see the review by \cite{Abraham01}. We therefore further restrict our redshift range to $z \leq 1$ in our full-sample analysis. Additionally, we increase our stellar mass limit to $\mathrm{log(M}_*/\mathrm{M}_{\odot}) = 9.5$ to reduce the effects of incompleteness (see Figure \ref{fig:mzsmplsel}) and to restrict our sample to a manageable number for morphological classifications. 
In Figure \ref{fig:mzsmplsel}, our final sample selection region is now shown as a solid rectangle. The distribution of the stellar mass and redshift of our final sample is also displayed in the right panels (b and c) of the same figure.  

Overall, our analysis of galaxies in different regions of the $M_*$-$z$ parameter space leads us to a final sample of galaxies for which we can confidently study their morphology and structure. We conclude that we can study the structure of galaxies up to $z \sim 1$ and down to log$(\mathrm{M}_{*}/\mathrm{M}_{\odot}) \geq 9.5$. Within this selection, our sample consists of $35,803$ galaxies with $14,036$ available spectroscopic redshifts.

\begin{figure}
\centering

  \centering
  \includegraphics[width=0.48\textwidth]{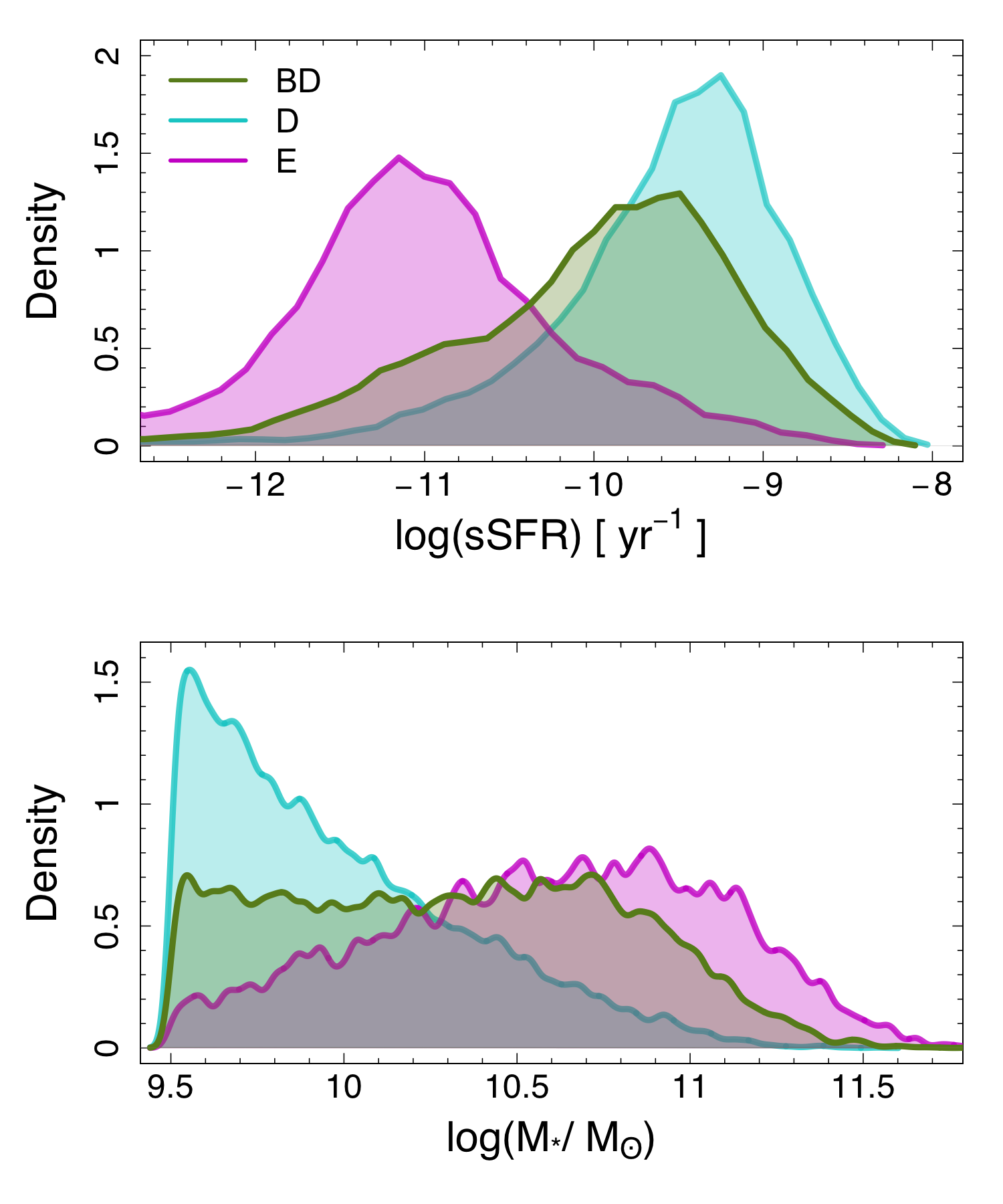}
  \label{subfig:stMass_PDF}
\caption{Top panel: the PDF of the specific star formation rate (sSFR = SFR/M$_*$) of the same morphologies. Bottom panel: the stellar mass probability density function (PDF) of three morphologies in our sample (all redshifts), i.e. \textit{pure-disk} (cyan), \textit{double-component} (green) and \textit{elliptical} (magenta) galaxies. Note that for clarity, the PDFs are slightly smoothed by a kernel with standard deviation of 0.02. }
\label{fig:stMass_sSFR_PDF}
\end{figure}

\section{Final Morphological Classifications} 
\label{sec:MorphClass}

In this section, we initially aim to develop a semi-automatic method for morphological classification. Our ultimate goal would be to reach a fully automatic algorithm for classifying galaxies into various morphological classes. While this ultimately proves unsuitable we discuss it here to explain why we finally visually inspect all systems.
These classes are as mentioned above \textit{bulge+disk} (BD; double-component), \textit{pure-disk} (D), \textit{elliptical} (E) and \textit{hard} (H) systems. In order to overcome this problem we test various methods including: cross-matching with Galaxy Zoo, Hubble catalogue (GZH) and Zurich Estimator of Structural Type (ZEST) catalogues. In the end, none of the methods proved to be robust and we resort to full visual inspection.      

\subsection{Using Galaxy Zoo: Hubble} 
\label{subsec:GZH}

A large fraction of COSMOS galaxies with I$_{F814W} < 23.5$ have been classified in the Galaxy Zoo: Hubble (GZH) project \citep{Willett17}. Like other Galaxy Zoo projects, this study classifies galaxies using crowdsourced visual inspections. They make use of images taken by the ACS in various wavelengths including 84,954 COSMOS galaxies in the F814W filter. Our sample has more than 80\% overlap with GZH, so we can cross-match and use their classifications to partially improve our final sample. We translate the GZH classifications into our desired morphological classes by using the decision tree shown in figure 4 of \cite{Willett17}, as well as the suggested thresholds for morphological selection presented in Table 11 of the same paper. Our final decision tree is displayed in the flowchart shown in Figure \ref{fig:GZH}. We refer to \cite{Willett17} for a detailed description of each task. The thresholds are shown as P values in the flowchart. In addition to using different combinations and permutations of tasks as shown in Figure \ref{fig:GZH}, the only part that we add to the GZH tasks is where an object is voted to have a smooth light profile. In this case, the object is likely to be an E or S0 or a smooth \textit{pure-disk} galaxy. To distinguish between them, we make use of the single S\'ersic index (n). As shown in the flowchart, $n > 2.5$ and $n \leq 2.5$ are classified as spheroid and \textit{pure-disk}, respectively. The S\'ersic indices are taken from our structural analysis which will be described in Hashemizadeh et al. (in prep.). In order to capture the S0 galaxies or \textit{double-component} systems with smooth disk profiles we add an extra condition as to whether there are at least 2 Galaxy Zoo votes for a prominent bulge. If so then it is likely that the galaxy is a \textit{double-component} system.
An advantage of using the GZH decision tree is that it can identify a vast majority of galaxies with odd features such as merger-induced asymmetry etc. 
Table \ref{tab:GZH_cofmat} shows a confusion matrix comparing the GZH predictions with the visual inspection of our 4k gold sample as the ground truth. Double component galaxies are predicted by the GZH with maximum accuracy 81\%. Single component (i.e. \textit{pure-disk} and \textit{elliptical} galaxies) and \textit{hard} galaxies are predicted at a significantly lower accuracy with high misclassification rates. We further visually inspected misclassified galaxies and do not find good agreement between our classification and those of GZH. As such, we do not trust the GZH classification for our sample.

\begin{figure*}
	\centering
	\includegraphics[width = \textwidth, height = 11.7cm]{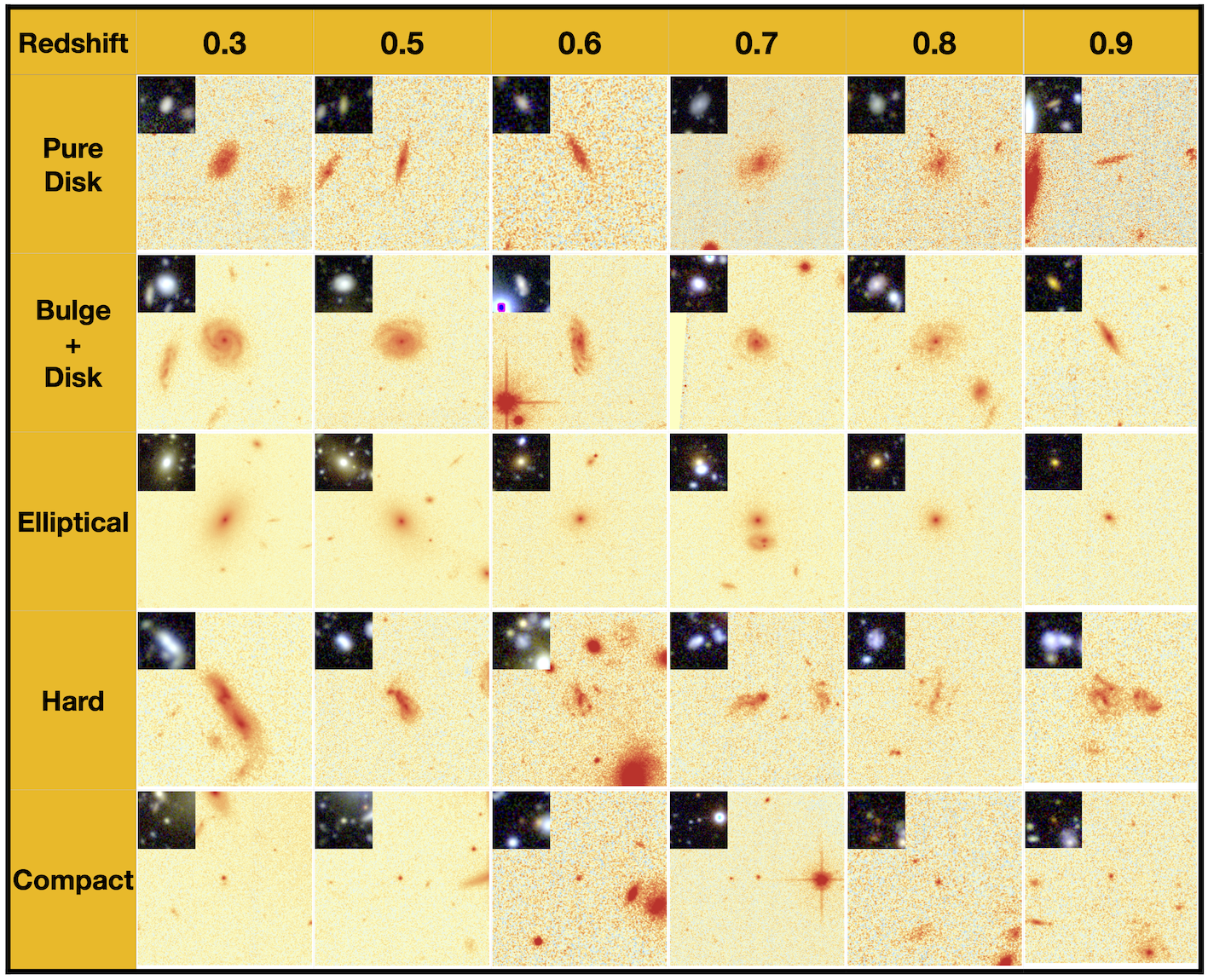}
	\caption{ Random postage stamps of various morphologies that we generate to perform our visual inspection. Background images are HST ACS/F814W while insets are combined SUBARU $gri$ colour image. Rows represent different morphologies. Galaxies are randomly selected within different redshift bins shown in columns. Redshifts annotated in the first row are the mean of the associated redshift bins. Cutouts indicate $5\times \mathrm{R90}$ on each side, where $\mathrm{R90}$ is measured from UltraVISTA Y-band in the DEVILS photometry catalogue (Davies et al. in prep.). }
	\label{fig:PostageStamp1}
\end{figure*}

\subsection{Using Zurich Estimator of Structural Type (ZEST)} 
\label{subsec:ZEST}

Another available morphological catalogue for COSMOS galaxies is the Zurich Estimator of Structural Type (ZEST) \citep{Scarlata07a}. In ZEST, \cite{Scarlata07a} use their single-S\'ersic index (n) as well as five other diagnostics: asymmetry, concentration, Gini coefficient, second-order moment of the brightest 20\% of galaxy flux, and ellipticity. ZEST includes a sample of $\sim 56,000$ galaxies with I$_{F814W}\leq 24$. More than 90\% of our sample is cross-matched with ZEST which we use as a complementary morphological classification. We use the ZEST TYPE flag with four values of 1,2,3,9 representing early type galaxy, disk galaxy, irregular galaxy and no classification. For Type 2 (i.e. disk galaxy) we make use of the additional flag BULG, which indicates the level of bulge prominence. BULG is flagged by five integers as follows: 0 = bulge dominated, 1,2 = intermediate-bulge, 3 = \textit{pure-disk} and 9 = no classification. 
We present the accuracy of ZEST predictions in a confusion matrix in Table \ref{tab:ZEST_cofmat} which shows we do not find an accurate morphological prediction from ZEST in comparison to our gold calibration sample. While \textit{double-component} systems are classified with an accuracy of 74\%, other classes are classified poorly with high error ratios. We confirm this by visual inspection of the misclassified objects where we still favour our visual classification.
\\[2\baselineskip]

Overall, by analysing both of the above catalogues we do not find them to be sufficiently accurate for predicting the proper morphologies of galaxies when we compare their estimates with our 4k gold calibration visual classification.   

\begin{table}
\centering
\caption{The confusion matrix comparing the morphological predictions of the GZH with our visual inspection of 4k gold calibration sample as the ground truth. For example, 0.81 means 81\% of double component galaxies (in our visual classification) are also correctly classified by the GZH.  }
\begin{tabular}{c|ccc}
\hline \hline
\backslashbox[35mm]{GZH Pred}{Ground Truth}
            & Double     & Hard          & Single \\ \hline \\
Double  & \textbf{0.81}  & 0.28          & 0.32         \\
Hard        & 0.0038         & \textbf{0.18} & 0.011      \\
Single      & 0.18           & 0.53 & \textbf{0.67}         \\ \\

\lasthline
\end{tabular}
\label{tab:GZH_cofmat}
\end{table}

\begin{table}
\centering
\caption{The confusion matrix comparing the morphological predictions of the ZEST catalogue with our visual inspection of 4k gold calibration sample as the ground truth. }
\begin{tabular}{c|ccc}
\hline \hline
\backslashbox[35mm]{ZEST Pred}{Ground Truth}
            & Double  & Hard & Single \\ \hline \\
Double  & \textbf{0.74}  & 0.34 & 0.42         \\
Hard        & 0.047  & \textbf{0.37} & 0.036       \\
Single      & 0.21  & 0.28 & \textbf{0.55}         \\ \\

\lasthline
\end{tabular}
\label{tab:ZEST_cofmat}
\end{table}

\subsection{Visual inspection of full D10/ACS sample} 
\label{subsec:VisInsp_full}

As no automatic classification robustly matches our gold calibration sample we opt to visually inspect all galaxies in our full sample. 
However, we can use the predictions from GZH and ZEST as a pre-classification decision and put galaxies in master directories according to their prediction. The \textit{hard} class is adopted from the GZH prediction as it is shown to perform well in detecting galaxies with odd features. In addition, from analysing the distribution of the half-light radius (R50) of our 4k gold calibration sample, extracted from our DEVILS photometric analysis, we know that resolving the structure of galaxies with a spatial size of R50 $\leq 0.15$ arcsec is nearly impossible. R50 is measured by using the {\sc ProFound} package \citep{Robotham18}, a tool written in the {\sc R} language for source finding and photometric analysis. So, we put these galaxies into a separate \textit{compact} class (C).

Having pre-classified galaxies, we now assign each class to one of our team members (AH, SPD, ASGR, LJMD, SB) so each of the authors is independently a specialist in, and responsible for, only one morphological class. Initially, we inspect galaxies and relocate incorrectly classified galaxies from our master directories to transition folders for further inspections by the associated responsible person. In the second iteration, the incorrectly classified systems will be reclassified and moved back into master directories. The left-over sources in the transition directories are therefore ambiguous. For these galaxies, all classifiers voted and we selected the most voted class as the final morphology. As a final assessment and quality check three of our authors (SPD, LJMD, SB) independently reviewed the entire classifications ($\sim 15k$ each) and identified $\sim 7k$ that were still felt to be questionable. These three classifiers then independently classified these 7k objects. The final classification was either the majority decision or in the case of a 3-way divergence the classification of SPD.

Figure \ref{fig:agreent} shows the stellar mass versus redshift plane colour coded according to the level of agreement between our 3 classifiers. Colour indicates the percentage of objects in each cell consistently classified by at least two classifiers, obtained as follows:

\begin{equation}
\mathrm{\scriptstyle Agreement} = \frac{\mathrm{\scriptstyle Number~of~objects~with~two~or~more~agreement~in~the~cell}}{\mathrm{\scriptstyle Total~number~of~objects~in~the~cell}},    
\label{eq:agree}
\end{equation}

This figure implies that we have the highest agreement ($\sim 100\%$) for high stellar mass objects at low redshifts. The agreement in our classification decreases to $\sim 90\%-95\%$ towards lower stellar mass regime at high redshift. This figure indicates that, on average, our visual classifications performed independently by different team members agree by $\sim95\%$ over the complete sample.

We report the number of objects in each morphological class in Table \ref{tab:FinalClassStats}. $44.4\%$ of our sample (15,931) are classified as \textit{double-component} (BD) systems. We find 13,212 \textit{pure-disk} (D) galaxies ($37\%$) while only 3,890 ($11\%$) \textit{elliptical} (E) galaxies. The \textit{Compact} (C) and \textit{hard} (H) systems, in total, occupy $7.6\%$ of our D10/ACS sample.

The visual morphological classification is released in the team-internal DEVILS data release one (DR1) in D10VisualMoprhologyCat data management unit (DMU). 

\subsection{Possible effects from the ``Morphological K-correction'' on our galaxy
classifications}
\label{subsec:K-corr}

The morphology of a subset of nearby galaxy classes shows some significant dependence on rest-frame wavelengths, especially below the Balmer break and towards the UV (e.g. \citealt{Hibbard97}; \citealt{Windhorst02}; \citealt{Papovich03}; \citealt{Taylor05}; \citealt{Taylor-Mager07}; \citealt{Huertas-Company09}; \citealt{Rawat09}; \citealt{Holwerda12}; \citealt{Mager18}). This ``Morphological K-correction'' can be quite significant, even between the B- and near-IR bands (e.g., \citealt{Knapen96}), and must be quantified in order to distinguish genuine evolutionary effects from simple bandpass shifting. Hence, the results of faint galaxy classifications may, to some extent, depend on the rest-frame wavelength sampled. 

Our D10/ACS classifications are done in the F814W filter (I-band), and the largest redshift in our sample, $z \sim$ 1, samples $\sim$412 nm (B-band), so for our particular case, the main question is, to what extent galaxy rest-frame morphology changes significantly from 412--823 nm across the BVRI filters. Here we briefly summarize how any such effects may affect our classifications. 

\cite{Windhorst02} imaged 37 nearby galaxies of all types with the Hubble Space Telescope (HST), gathering available data mostly at 150, 255, 300, 336, 450, 555, 680, and/or 814 nm, including some ground based images to complement filters missing with HST. These nearby galaxies are all large enough that a ground-based V-band image yields the same classification as an HST F555W
or F606W image. These authors conclude that the change in rest-frame morphology going from the red to the mid--far UV is more pronounced for early type galaxies (as defined at the traditional optical wavelengths or V-band), but not necessarily negligible for all mid-type spirals or star-forming galaxies. Late-type galaxies generally look more similar in morphology from the mid-UV to the red due to their
more uniform and pervasive SF that shows similar structure for young--old stars in all filters from the mid-UV through the red. \cite{Windhorst02} conclude {\it qualitatively} that in the rest-frame mid-UV, early- to mid-type galaxies are more likely to be misclassified as later types than late-type galaxies are likely to be misclassified as earlier types. 

To {\it quantify} these earlier qualitative findings regarding the morphological K-correction, much larger sample of 199 nearby galaxies across all Hubble types (as defined in V-band) was observed by \cite{Taylor05} and \cite{Taylor-Mager07}, and 2071 nearby galaxies were similarly analyzed by \cite{Mager18}. They determined their SB-profiles, radial light-profiles, color gradients, and CAS parameters (Concentration index, Asymmetry, and Clumpiness; e.g. \citealt{Conselice04}) as a function of rest-frame wavelength from 150-823 nm.

\cite{Taylor-Mager07} and \cite{Mager18} conclude that early-type galaxies (E--S0) have CAS parameters that appear, within their errors, to be similar at all wavelengths longwards of the Balmer break, but that in the far-UV, E--S0 galaxies have concentrations and asymmetries that more closely resemble those of spiral and peculiar/merging galaxies in the optical. This may be explained by extended disks containing recent star formation. The CAS parameters for galaxy types later than S0 show a mild but significant wavelength dependence, even over the wavelength range 436-823 nm, and a little more significantly so for the earlier spiral galaxy types (Sa--Sc). These galaxies generally become less concentrated and more asymmetric and somewhat more clumpy towards shorter wavelengths. The same is true for mergers when they progress from pre-merger via major-merger to merger-remnant stages.

While these trends are mostly small and within the statistical error bars for most galaxy types from 436--823 nm, this is not the case for the Concentration index and Asymmetry of Sa--Sc galaxies. For these galaxies, the Concentration
index decreases and the Asymmetry increases going from 823 nm to 436 nm (Fig. 17 of \citealt{Taylor-Mager07} and Fig. 5 of \citealt{Mager18}). Hence, to the extent that our visual classifications of apparent Sa--Sc galaxies depended on their Concentration index and Asymmetry, it is possible that some of these objects may have been misclassified. E-S0 galaxies show no such trend in Concentration with wavelength for 436--823 nm, and have Concentration indices much higher than Sa--Sc galaxies and Asymmetry and Clumpiness parameters generally much lower than Sa--Sc galaxies. Hence, it is not likely that a significant fraction of E--S0 galaxies are misclassified as Sa--Sc galaxies at $z \lesssim 0.8$ in the 436--823 nm filters due to the Morphological K-correction.

Sd-Im galaxies show milder trends in their CAS parameters and in the same direction as the Sa--Sc galaxies. It is thus possible that a small fraction of Sa-Sc galaxies --- if visually relying heavily on Concentration and Asymmetry index --- may have been visually misclassified as Sd--Im galaxies, while a smaller fraction of Sd--Im galaxies may have been misclassified as Sa--Sc galaxies. The available data on the rest-frame wavelength dependence of the morphology of nearby galaxies does not currently permit us to make more precise statements. In any case, these CAS trends with restframe wavelengths as observed at 436--823 nm for redshifts $z\simeq$ 0.8--0 in the F814W filter are shallow enough that the fraction of misclassifications between Sa--Sc and Sd--Im galaxies, and vice versa, is likely small. 

\subsection{Review} 
\label{subsec:VisInsp_verify}

In Figure \ref{fig:frac_z1}, we show the associated fractions of our final visual inspection as solid lines. The global trends are in good agreement with our initial 4k gold calibration sample. 
We find that, although our inspection procedure may have slightly changed from the 4k gold calibration sample to the full sample, the outcome classifications are consistent in the three primary classes making up more than $92\%$ of our sample.
 
 \begin{figure*}
	\centering
	\includegraphics[width=\textwidth]{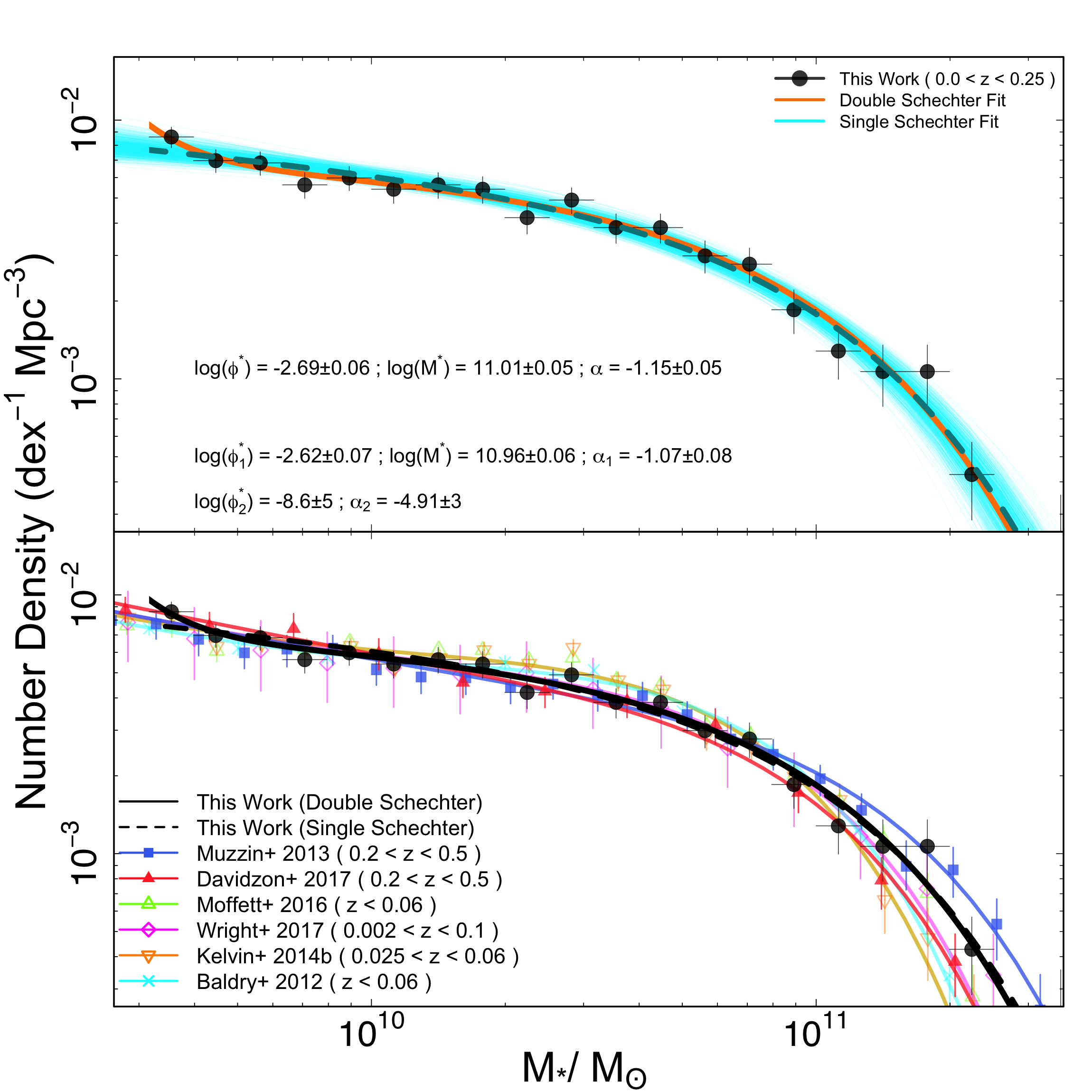}
	\caption{Top panel: Single and double Schechter functions fitted to a low-$z$ sample of D10/ACS ($0.0 < z < 0.25$). The fits are performed by using {\fontfamily{qcr}\selectfont dftools} down to the stellar mass limit of the sample, i.e. $10^{9.5} M_\odot$. 
    Transparent region shows the error range calculated by 1000 times sampling of the full posterior probability distribution of the single Schechter fit parameters. Black data points show the binned galaxy counts for which the stellar mass range from minimum to maximum ($9.5 \le$ log$(\mathrm{M}_{*}/\mathrm{M}_{\odot}) \le 12$) is binned into 25 equal-sized bins. Note that the Schechter functions are not fitted directly to the galaxy counts. Bottom panel: Comparison of our SMF at $0.0 < z < 0.25$ with \protect\cite{Muzzin13} and \protect\cite{Davidzon17}. For completeness, the SMF of the GAMA local galaxies from several studies are shown in different colours. Solid and dashed black lines represent the double and single Schechter function fits to our data, respectively (eq. \ref{eq:SingleSchechter} \& \ref{eq:DoubleSchechter}). Note that, the fits to other studies (colour solid lines) are double Schechter fits.}
	\label{fig:Mfunc_z0}
\end{figure*}

\begin{table}
\centering
\caption{Final number of objects in each of our morphological classes. }
\begin{tabular}{ccc}
\firsthline \firsthline 
Morphology  & Object Number  & Percentage \\ \hline \\
Double  & 15,931  & 44.4\%  \\
Pure Disk & 13,212 & 37\%     \\
Elliptical & 3,890 & 11\%      \\
Compact & 1,124 & 3.1\%       \\
Hard & 1,600 & 4.5\%          \\ \\

\lasthline
\end{tabular}
\label{tab:FinalClassStats}
\end{table}

Figure \ref{fig:stMass_sSFR_PDF} shows the probability density function (PDF) of the sSFR (upper panel) and total stellar mass (lower panel) for galaxies classified as D, BD and E. We use the SFR from the {\sc ProSpect} SED fits described in \cite{Thorne20}. These figures indicate that, as one would expect, D galaxies dominate lower stellar mass and higher sSFR regime, opposite to Es which occupy the high stellar mass end and low sSFR. BD galaxies are systems located in-between these two classes in terms of both stellar mass and sSFR. These results are as expected and provide some confidence that our classifications are sensible.

Figure \ref{fig:PostageStamp1} displays a set of random galaxies in each of the morphological classes within different redshift intervals. In addition, we present 50 random galaxies of each morphology in Figures \ref{fig:contact_sheets_double}-\ref{fig:contact_sheets_comp}. 

Having finalised our morphological classification, we now investigate the stellar mass functions for different morphologies and their evolution from $z = 1$.

\section{Fitting galaxy stellar mass function} 
\label{sec:FitSMF}

For parameterizing the SMF, we assume a fitting function that can describe the galaxy number density, $\Phi\left(\mathrm{M}\right)$. The typical function adopted is that described by \cite{Schechter76} as:  

\begin{equation}
\Phi(M)\mathrm{d}M=\Phi^{*} e^{-M / M^{*}}\left(\frac{M}{M^{*}}\right)^{\alpha} \mathrm{d}M,
\label{eq:SingleSchechter}
\end{equation}

\noindent where the three key parameters of the function are, $\alpha$, the power low slope or the slope of the low-mass end, $\Phi^{*}$, the normalization, and, $M^{*}$, the characteristic mass (also known as the break mass or the knee of the Schechter function). 

At very low redshifts a number of studies have argued that the shape of the SMF is better described by a double Schechter function (\citealt{Baldry08}; \citealt{Pozzetti10}; \citealt{Baldry12}), i.e. a combination of two single Schechter functions, parameterized by a single break mass ($M^{*}$), and given by: 

\begin{equation}
\Phi(M) \mathrm{d} M=e^{-M / M^{*}}\left[\Phi_{1}^{*}\left(\frac{M}{M^{*}}\right)^{\alpha_{1}}+\Phi_{2}^{*}\left(\frac{M}{M^{*}}\right)^{\alpha_{2}}\right] \frac{\mathrm{d} M}{M^{*}},
\label{eq:DoubleSchechter}
\end{equation}

\noindent where $\alpha_2 < \alpha_1$, indicating that the second term predominantly drives the lower stellar mass range.

To fit our SMFs, we use a modified maximum likelihood (MML) method (i.e., not $1/V_{\mathrm{max}}$) as implemented in {\fontfamily{qcr}\selectfont dftools}\footnote{\href{https://github.com/obreschkow/dftools}{https://github.com/obreschkow/dftools}} \citep{Obreschkow18}. This technique has multiple advantages including: it is free of binning, accounts for small statistics and Eddington bias. {\fontfamily{qcr}\selectfont dftools} recovers the mass function (MF) while simultaneously handling any complex selection function, Eddington bias and the cosmic large-scale structure (LSS). Eddington bias tends to change the distribution of galaxies, particularly in the low-mass regime which is more sensitive to the survey depth and S/N, as well as high-mass regime due to the exponential cut-off which is sensitive to the scatters by noise (e.g. \citealt{Ilbert13}; \citealt{Caputi15}). Eddington bias occurs because of the observational/photometric errors and often can dominate over the shot noise \citep{Obreschkow18}. Photometric uncertainties, which are introduced in the redshift estimation, as well as the stellar mass measurements are fundamentally the source of this bias (\citealt{Davidzon17}). We account for Eddington bias in {\fontfamily{qcr}\selectfont dftools} by providing the errors on the stellar masses from the {\sc ProSpect} analysis by \cite{Thorne20}.

The R language implementation and MML method make {\fontfamily{qcr}\selectfont dftools} very fast. In the fitting procedures described in this work, we use the inbuilt {\fontfamily{qcr}\selectfont optim} function with the default optimization algorithm of \cite{Nelder65} for maximizing the likelihood function. In order to account for the volume corrections, we use the effective unmasked area of the D10/ACS region which we calculate to be $1.3467$ square degrees. This is exclusive of the masked areas from bright stars and the non-uniform edges of the ACS mosaic (see Figure \ref{fig:F814W_DEVILS_smp_z1}) and this is calculated using the process outlined in Davies et al. (in prep.). Our selection function, required for a proper volume corrected distribution function, is essentially a volume limited sample, i.e. a constant volume across adopted mass range. We refer the reader to \cite{Obreschkow18} for full details regarding {\fontfamily{qcr}\selectfont dftools} and its methodology. In this paper, we fit both single and double Schechter functions to the SMF. We examine both functions and will discuss further in Section \ref{subsec:MfunZ0}.

\subsection{Verification of our SMF fitting process at low redshift} 
\label{subsec:MfunZ0}  

We first validate our \textit{total} SMF fitting process at low-$z$ and compare with the known literature, prior to splitting by redshift and morphology.

To achieve this, we compare the SMF of D10/ACS galaxies in our lowest-$z$ bin, i.e. $0 < z < 0.25$, with literature studies. We primarily choose this redshift range to compare with the SMF \cite{Muzzin13} and \protect\cite{Davidzon17} at $0.2 < z < 0.5$ in the COSMOS/UltraVISTA field and local GAMA galaxies at $z < 0.06$ (\citealt{Baldry12}; \citealt{Kelvin14}; \citealt{Moffett16a}; \citealt{Wright17}). We fit the SMF within this redshift range using both single and double Schechter functions (Equations \ref{eq:SingleSchechter} and \ref{eq:DoubleSchechter}, respectively). The upper panel of Figure \ref{fig:Mfunc_z0}, shows our single and double Schechter fit to this low-$z$ sample. As annotated in the figure, we report the best fit single Schechter parameters of log$(\mathrm{M}^{*}/\mathrm{M}_{\odot}) = 11.01\pm0.05$, $\alpha = -1.15\pm0.05$, log$(\Phi^*) = -2.69\pm0.06$ and log$(\mathrm{M}^{*}/\mathrm{M}_{\odot}) = 10.96\pm0.06$ $\alpha_1 = -1.07\pm0.08$, $\alpha_2 = -4.91\pm3$, log$(\Phi^*_1) = -2.62\pm0.07$ and log$(\Phi^*_2) = -8.6\pm5$ in our double Schechter fit. Therefore, the double Schechter fit involves a broad range of uncertainty, in particular at the low-mass range. 
Larger uncertainty is likely due to the fact that the stellar mass range of the D10/ACS sample does not extend to log$(M_*/M_\odot) < 9.5$, where the pronounced upturn in the SMF occurs. 
The error ranges shown in the upper panel of Figure \ref{fig:Mfunc_z0} as transparent curves are obtained from 1000 samples drawn from the full posterior probability distribution of all the single Schechter parameters. 

Despite larger errors on the double Schechter parameters, we elect to use this function for our future analysis at all redshifts as it has been shown that a double Schechter can better describe the stellar mass distribution even at higher-$z$ (e.g., \citealt{Wright18}).

In the lower panel of Figure \ref{fig:Mfunc_z0}, we show that our SMF is in good agreement with \cite{Muzzin13} and \protect\cite{Davidzon17} within the quoted errors.
Overall, we see a good agreement with the mass function of galaxies at low-$z$. We observe no significant flattening of the SMF at the intermediate stellar masses ($10^{9.5}-10^{10.5}$) as reported in the local Universe by e.g., \cite{Moffett16a}. We are unsurprised that we do not see this flattening as the D10/ACS contains only 3 galaxies in the redshift range comparable to GAMA ($z < 0.06$). 

One might expect an evolution of the SMF within $0 < z < 0.25$ that would explain the slightly higher number density in the intermediate mass range. Our fitted Schechter functions however also deviate from the GAMA SMFs in the high stellar mass end. We find that this is systematically due to the Schechter function fitting process. Unlike D10/ACS, as shown in Figure \ref{fig:Mfunc_z0}, the local literature data extend to lower stellar mass regimes (log$(\mathrm{M}_{*}/\mathrm{M}_{\odot}) = 8$ and $7.5$ in the case of \citealt{Wright17}), influencing the bright end fit and the position of $\mathrm{M}^*$. In this regime the upturn of the stellar mass distribution is remarkably more pronounced. This strong upturn drives the optimization fitting and impacts the high-mass end. This highlights the difficulty in directly comparing fitted Schechter values if fitted over different mass ranges. 

\subsection{Cosmic Large Scale Structure Correction} 
\label{subsec:LSS_cor}

\begin{figure}
	\centering
	\includegraphics[width=0.49\textwidth]{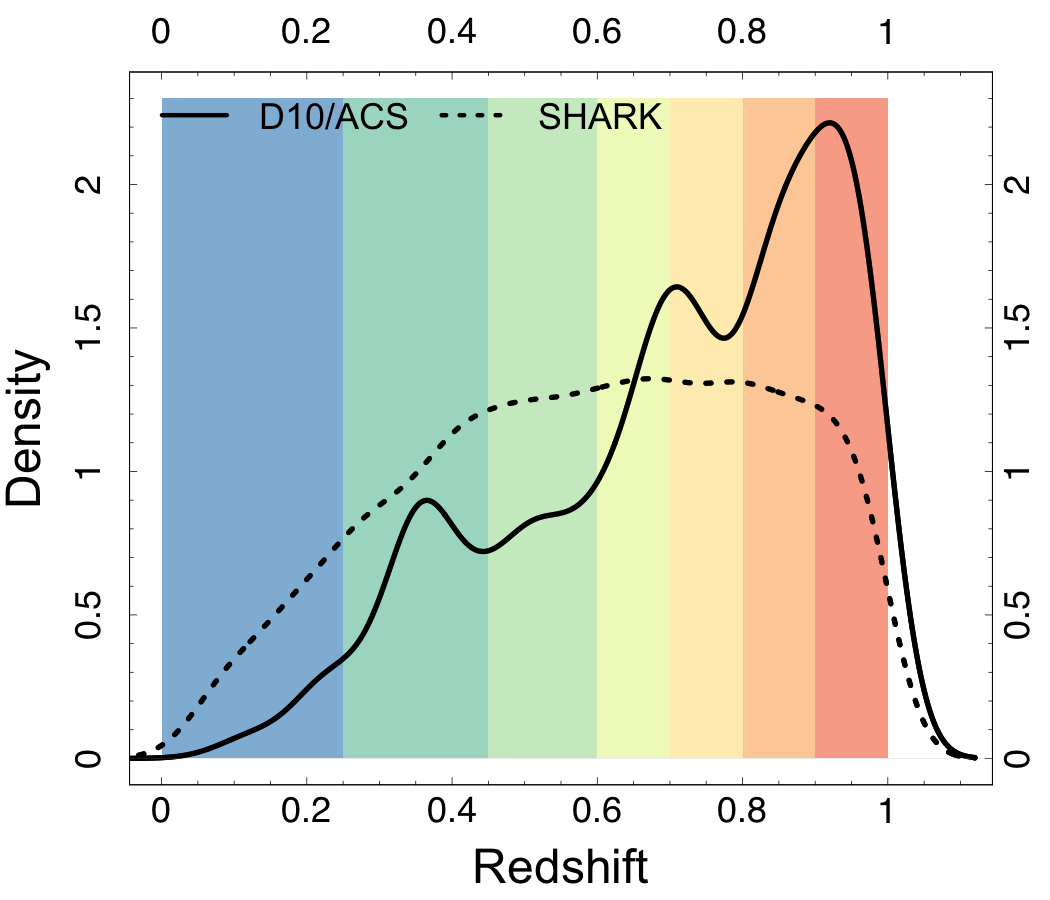}
	\caption{ The $\mathrm{N}(z)$ distribution of the D10/ACS sample (solid line) compared with the SHARK semi-analytic model prediction (dashed line). SHARK data represent a light-cone covering 107.889 square degrees with Y-mag $< 23.5$.  Colour bands represent the redshift bins that we consider in this work. Note that the PDFs are smoothed by a kernel with standard deviation of 0.3 so are non-zero beyond the nominal limits.}
	\label{fig:zdens}
\end{figure}

\begin{figure}
	\centering
	\includegraphics[width = 0.49\textwidth, angle = 0]{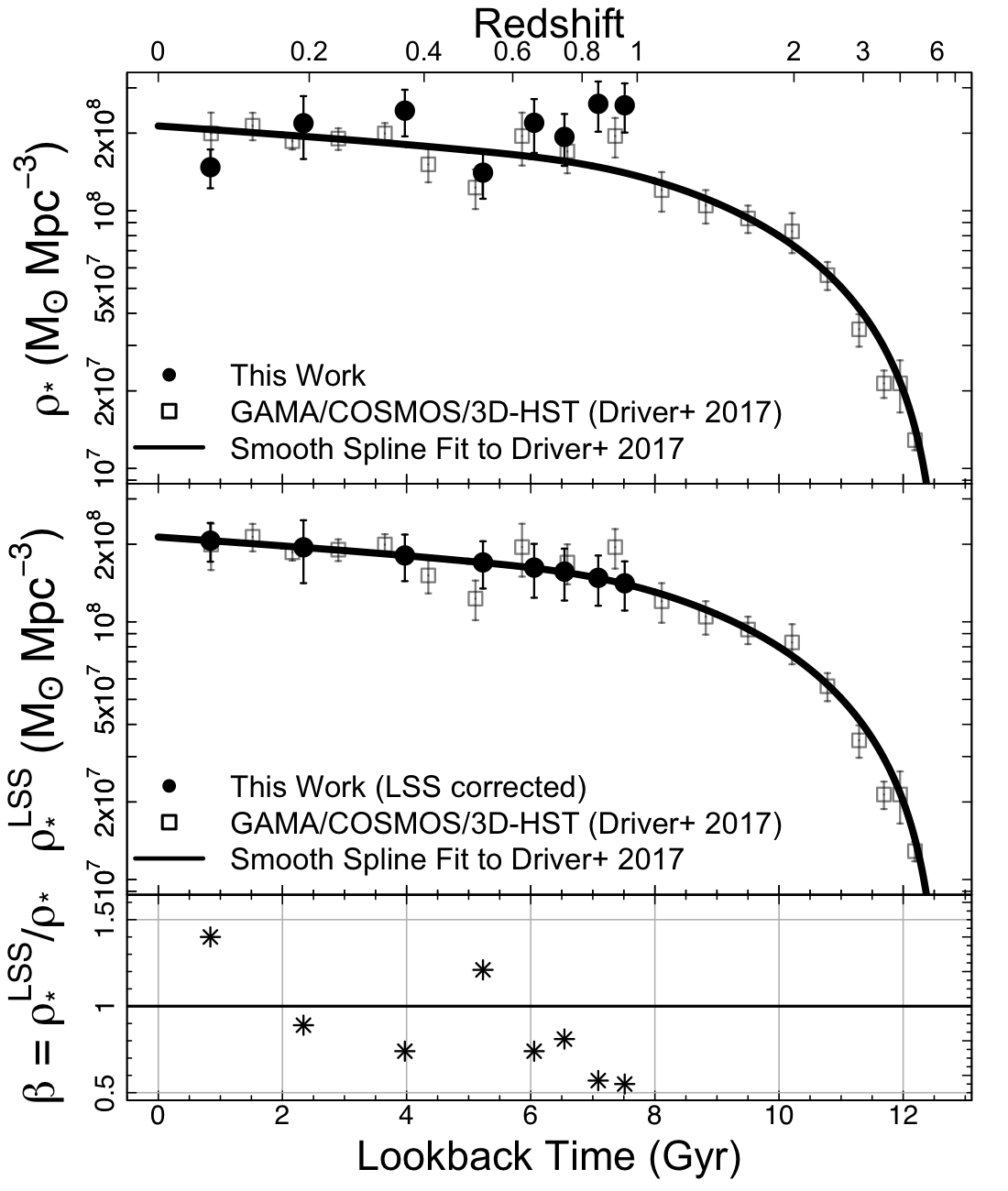}
	\caption{ Top panel: our measurement of the evolution of the \textit{total} SMD (per comoving Mpc$^{3}$) compared with a compilation of GAMA, COSMOS and 3D-HST by \protect\cite{Driver18}. The black curve represents a spline fit to the \protect\cite{Driver18} data. Middle panel: the large scale structure correction factor applied to our SMD in order to meet the predictions of the spline fit. Bottom panel: the residual of the SMDs before and after the LSS correction indicating the correction coefficient we apply to each redshift bin. }
\label{fig:LSS_cor}
\end{figure}

All galaxy surveys are to some extent influenced by the cosmic large scale structure (LSS, \citealt{Obreschkow18}). Generally, the LSS produces local over- and under-densities of the galaxies at particular redshifts in comparison to the mean density of the Universe at that epoch. For example, GAMA regions are $\sim 10\%$ underdense compared with SDSS (\citealt{Driver11}).  

We observe this phenomenon in the nonuniform D10/ACS redshift distribution, $\mathrm{N}(z)$.
To highlight this, Figure \ref{fig:zdens} compares the $\mathrm{N}(z)$ distribution of the D10/ACS sample with the prediction of the SHARK semi-analytic model (\citealt{Lagos18b}; \citealt{Lagos19}). SHARK data in this figure represent a light-cone covering 107.889 square degrees with Y-mag $< 23.5$, and because of the much larger simulated volume, is less susceptible to LSS. In this figure the redshift bins we shall use later in our analysis are shown as background colour bands indicating redshift intervals of (0, 0.25, 0.45, 0.6, 0.7, 0.8, 0.9 and 1.0). SHARK predicts a nearly uniform galaxy distribution with no significant over- and under-density regions, while the empirical D10/ACS sample shows a nonuniform $\mathrm{N}(z)$ with overdensities and underdensities.
These density fluctuations can introduce systematic errors in the construction of the SMF by, for example, overestimating the number density of very low-mass galaxies which are only detectable at lower redshifts \citep{Obreschkow18}. In other words, the LSS can artificially change the shape/normalization of the SMF.

Using the distance distribution of galaxies, {\fontfamily{qcr}\selectfont dftools} (\citealt{Obreschkow18}) internally accounts for the LSS by modeling the relative density in the survey volume, $g(r)$, i.e. it measures the mean galaxy density of the survey at the comoving distance $r$, relative to the mean density of the Universe. Incorporating this modification into the effective volume, the MML formalism works well for a sensitivity-limited sample (see \citealt{Obreschkow18} for details). However, for our volume limited sample, this method is unable to thoroughly model the density fluctuations. Therefore, to take this non-uniformity into account, we perform a manual correction as follows.

In a comprehensive study of the cosmic star formation history, \cite{Driver18} compiled GAMA, COSMOS and 3D-HST data to estimate the total stellar mass density from high redshifts to the local Universe ($0 < z < 5$). Figure \ref{fig:LSS_cor} (upper panel) shows these data. As we know the total stellar mass density must grow smoothly and hence we assume that perturbations around a smooth fit represent the underlying LSS. 
In Figure \ref{fig:LSS_cor} (middle panel), we fit a smooth spline with the degree of freedom $3.5$ to the \cite{Driver18} data to determine a uniform evolution of the total $\rho_*$. We then introduce a set of correction factors ($\beta$) to our empirical measurements of the \textit{total} SMDs to match the prediction of the spline fit (middle panel of Figure \ref{fig:LSS_cor}). We then apply these correction factors (bottom panel of Figure \ref{fig:LSS_cor}) to our estimations of the morphological SMD values.
This ignores any coupling between morphology and LSS which we consider a second order effect.
We report the $\beta$ correction factors in Table \ref{tab:rho} and show them in the bottom panel of Figure \ref{fig:LSS_cor}.

\begin{landscape}
\begin{figure}
	\centering
	\includegraphics[width = 1.3\textwidth, angle = 0]{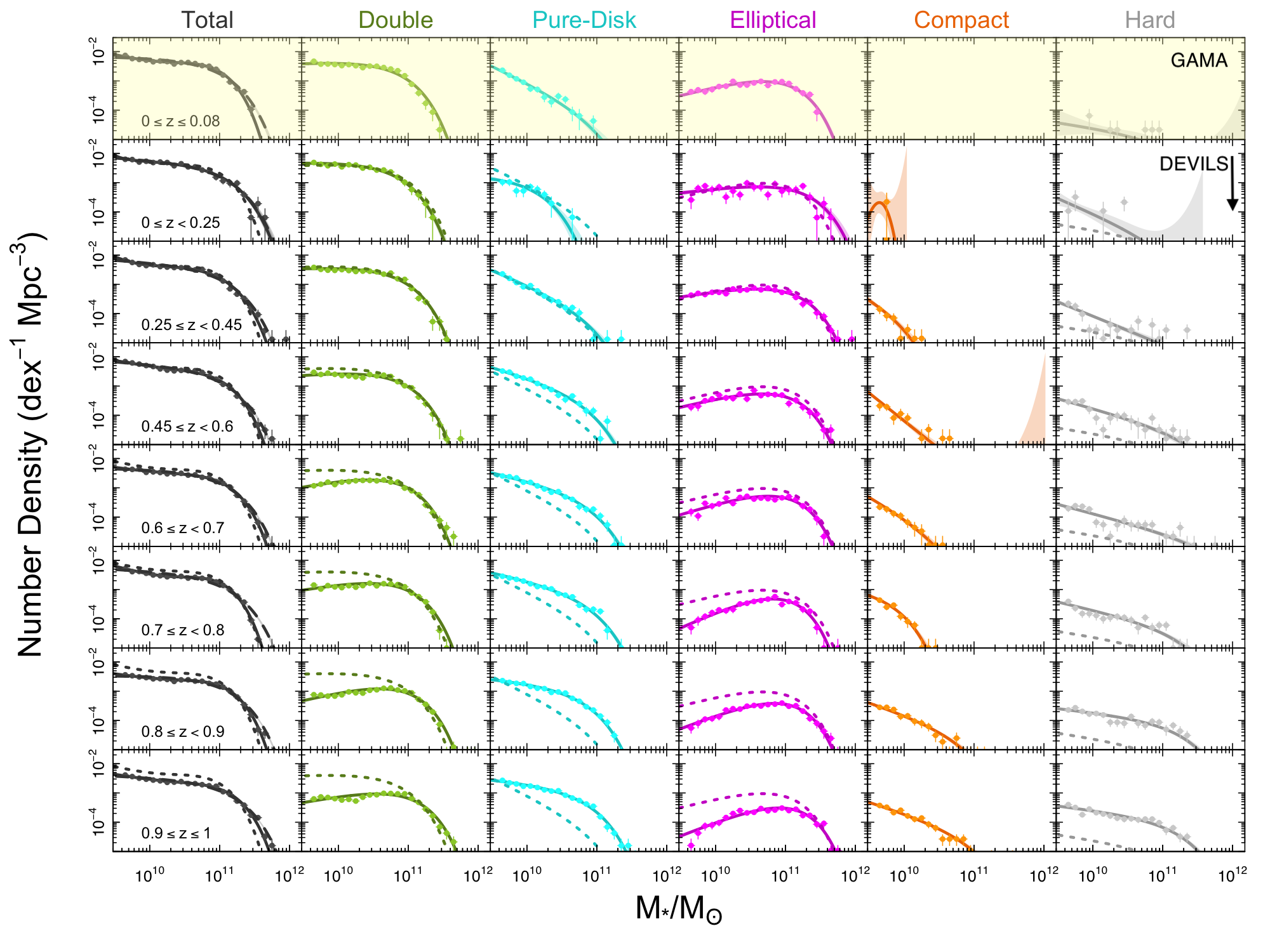}
	\caption{ The \textit{total} and morphological SMFs in eight redshift bins. Top row highlighted by yellow colour represents the GAMA SMFs ($0 \le z \le 0.08$). Data points are galaxy counts in each of equal-size stellar mass bins. Width of stellar mass bins are shown as horizontal bars on data points. Vertical bars are poisson errors. Shaded regions around the best fit curves are 68 per cent confidence regions. Black solid and dashed curves demonstrate double and single Schechter functions of all galaxies, respectively, while dotted curves over-plotted on higher-$z$ bins are the GAMA $z=0$ SMFs to highlight the evolution of the SMF.}
	\label{fig:Mfunc_6z}
\end{figure}
\end{landscape}

\begin{figure*}
	\centering
	\includegraphics[width=\textwidth]{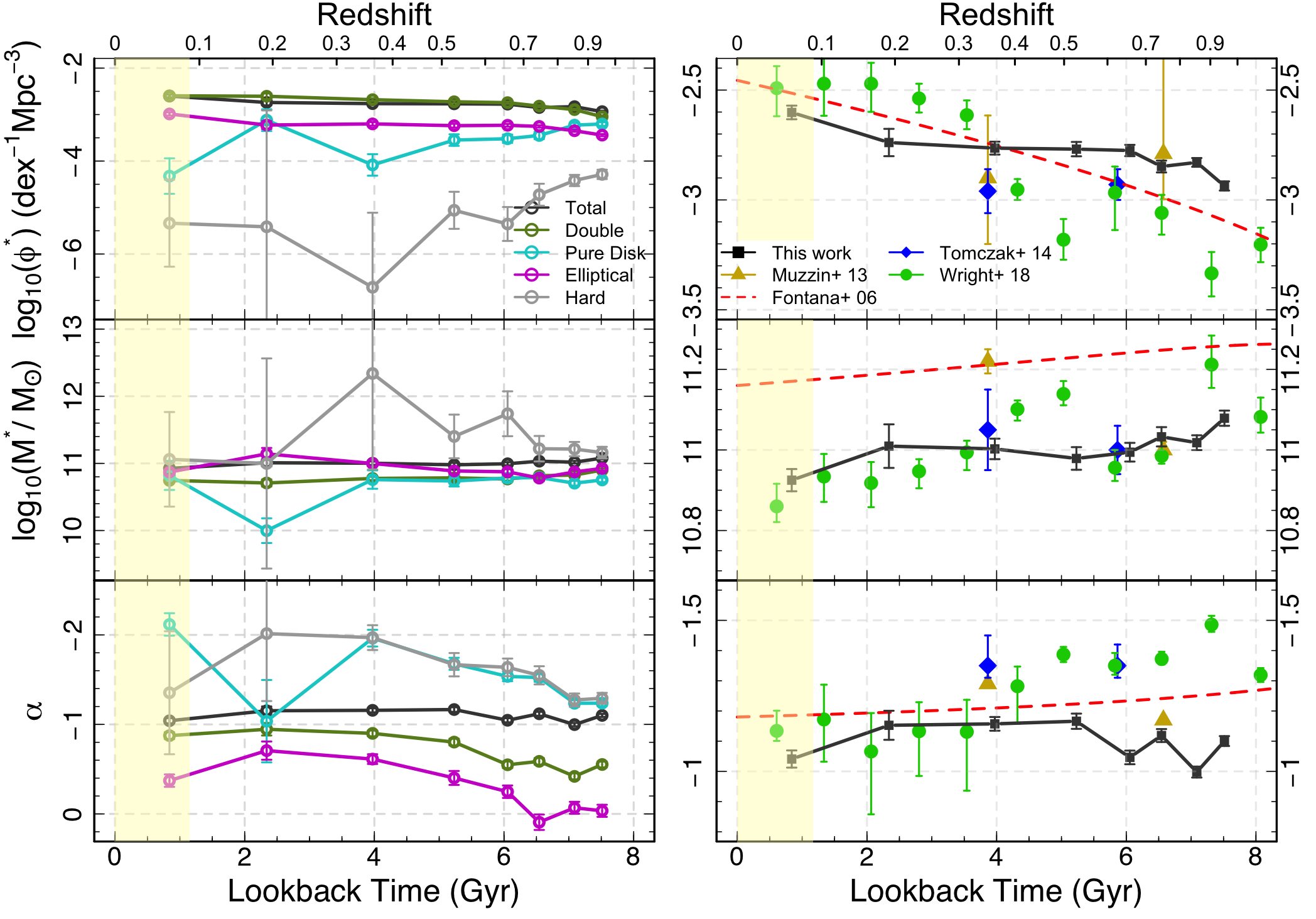}
	\caption{ Left: Evolution of the single Schechter best fit parameters, $\Phi^*$, $M^*$ and $\alpha$. Error bars are the standard deviations on each of the most likely parameters. Black line is the \textit{total} SMF while colour coded data represent the best fit parameters of the Schechter function of different morphologies. Note that for simplicity we remove \textit{compacts} as their trends are very noisy and washes out the trends of other morphological types. Right: Zoomed-in plot showing the evolution of our \textit{total} Schechter function parameters compared with a compilation of other studies. Highlighted region shows the epoch covered by the GAMA data.}
	\label{fig:Mfunc_6z_par_evol}
\end{figure*}

\subsection{Evolution of the SMF since \lowercase{$z$} $= 1$} 
\label{subsec:MfunEvol}

With the LSS correction in place, we now split the sample into 7 bins of redshift (0-0.25, 0.25-0.45, 0.45-0.6, 0.6-0.7, 0.7-0.8, 0.8-0.9 and 0.9-1.0) so that we have enough numbers of objects in each bin and are not too wide to incorporate a significant evolution (these redshift ranges are shown by the colour bands in Figure \ref{fig:zdens}). Each bin contain 1,108, 5,041, 4,216, 4,867, 5,277, 7,090, 8,207 galaxies, respectively.

Figure \ref{fig:Mfunc_6z} shows the SMF in each redshift bin for the full sample, as well as for different morphological types of: \textit{double-component} (BD), \textit{pure-disk} (D), \textit{elliptical} (E), \textit{compact} (C) and \textit{hard} (H). These SMF measurements include the correction for LSS as discussed in Section \ref{subsec:LSS_cor}. Note that the first row of Figure \ref{fig:Mfunc_6z} highlighted by yellow shade shows GAMA $z = 0$ SMFs (Driver et al., in prep.).

As shown in Figure \ref{fig:Mfunc_6z}, we fit the \textit{total} SMF within all 8 redshift bins (including GAMA) by both single and double Schechter functions (displayed as dashed and solid black curves, respectively), while the morphological SMFs are well fit by single Schechter function at all epochs. The difference between double and single Schechter fits is insignificant, compared to the error on individual points. However, as noted in Section \ref{subsec:MfunZ0}, for calculating the stellar mass density we will use our double Schechter fits.
As we will see later, the effect of this choice on the measurement of our total stellar mass density is negligible. In Figure \ref{fig:Mfunc_6z}, we over-plot the GAMA $z = 0$ SMFs (Driver et al., in prep.) on higher-$z$ to highlight the evolution (dotted curves). 
In the \textit{total} SMF we see a growth at the low-mass end and a relative stability at high-mass end, particularly when comparing high-$z$ with the D10/ACS low-$z$ ($0 < z < 0.25$). 
At face value, this suggests that since $z = 1$ in-situ star formation and minor mergers (low-mass end) play an important role in forming or transforming mass within galaxies \citep{Robotham14}. 

Looking at the SMF of the morphological sub-classes we find that the BD systems show no significant growth with time at their high-mass end, while at the low-mass end we see a noticeable increase in number density. 
In D galaxies, at both low- and high-mass ends we find a variation in the number density with the low-mass end increasing and extreme steepening at the lower-$z$ and high-mass end decreasing with time. Note that we do not rule out the possible effects of some degree of incompleteness in this mass regime on the evolution of the low-mass end. 
For E galaxies, however, we report a modest growth in their high-mass end (again when comparing with D10/ACS low-$z$) and a significant growth in intermediate- and low-mass regimes from $z = 1$ to $z = 0$. 
Finally, H and C systems become less prominent with declining redshift as fewer galaxies occupy these classes. The physical implications of these trends will be discussed in Section \ref{sec:discussion}.  

Figure \ref{fig:Mfunc_6z_par_evol} further summarizes the trends in Figure \ref{fig:Mfunc_6z}, showing the evolution of our best fit single Schechter parameters $\Phi^*$, $M^*$ and $\alpha$. We report the best Schechter fit parameters in Table \ref{tab:Morph_MF_par}. For comparison, we also show a compilation of literature values in the right panel. The literature data show only single Schechter parameters of the \textit{total} SMF. Comparing our \textit{total} SMF with other studies, including \cite{Muzzin13}, we find a good agreement within the quoted errors. 

Note that the three parameters are, of course, highly correlated and the mass limits to which they are fitted vary.    

The $\Phi^*$ of BD systems has the largest value and almost mimics the trend of the \textit{total} $\Phi^*$ which is largely consistent with no evolution. This is expected as BD galaxies dominate the sample at almost all redshifts. We report a slight decrease in the $\Phi^*$ value of D galaxies, while almost no evolution in E systems. Note that as mentioned earlier, our lowest redshift bin ($0.0 < z < 0.25$) contains only $1,108$ galaxies of which only $102$ are D systems incorporating an uncertainty in our fitting process. This can be seen in the deviation of the second data point of D galaxies (cyan lines) from other redshifts in Figure \ref{fig:Mfunc_6z_par_evol}.  
H systems also contribute more at higher redshifts (higher $\Phi^*$ values) owing to the fact that the abundance of mergers, clumpy and disrupted systems dramatically increases at higher redshifts (despite lower resolution). We note that due to our LSS corrections (Section \ref{subsec:LSS_cor}) the $\Phi^*$ reported here is not the directly measured parameter but modified according to our LSS correction coefficients (reported in Table \ref{tab:rho}). This normalization smooths out the evolutionary trends, otherwise fluctuating due to the large scale structures, but does not impact their global trends.

The characteristic mass, $M^*$, of the BD systems presents a stable trend since $z~=~1$. We interpret this behaviour to be a result of the lack of significant evolution of the massive/intermediate-mass regime of the SMF of this morphological class. D galaxies also show slight evidence of evolution with some fluctuations in $M^*$ value. E galaxies, likewise, evolve only a small amount. Overall, we observe a behaviour consistent with no evolution in $M^*$ for all morphological types. Note that large variation of the $M^*$ in H and C types, at low-$z$ in particular, is not physical. This is massively driven by the lack of H and C galaxies at low-$z$ resulting in an unconstrained turnover and as can be seen in Figure \ref{fig:Mfunc_6z_par_evol} large uncertainties.

The low-mass end slope, $\alpha$, of different morphological classes also shows some evolution. Similar to D systems, BD galaxies show a marked steepening increase in their slope at later times, indicating that the SMF steepens at lower redshifts. The steepest mass function at almost all times is for the D galaxies. E galaxies occupy the lowest steepening but constantly growing from $\alpha = 0.34$ to $-0.75$ at $z \sim 0.2$.

\section{The Evolution of the Stellar Mass Density Since \lowercase{$z$} $= 1$} 
\label{sec:rho}

\begin{landscape}
 \begin{figure}
	\centering
	\includegraphics[width = 1.3\textwidth, angle = 0]{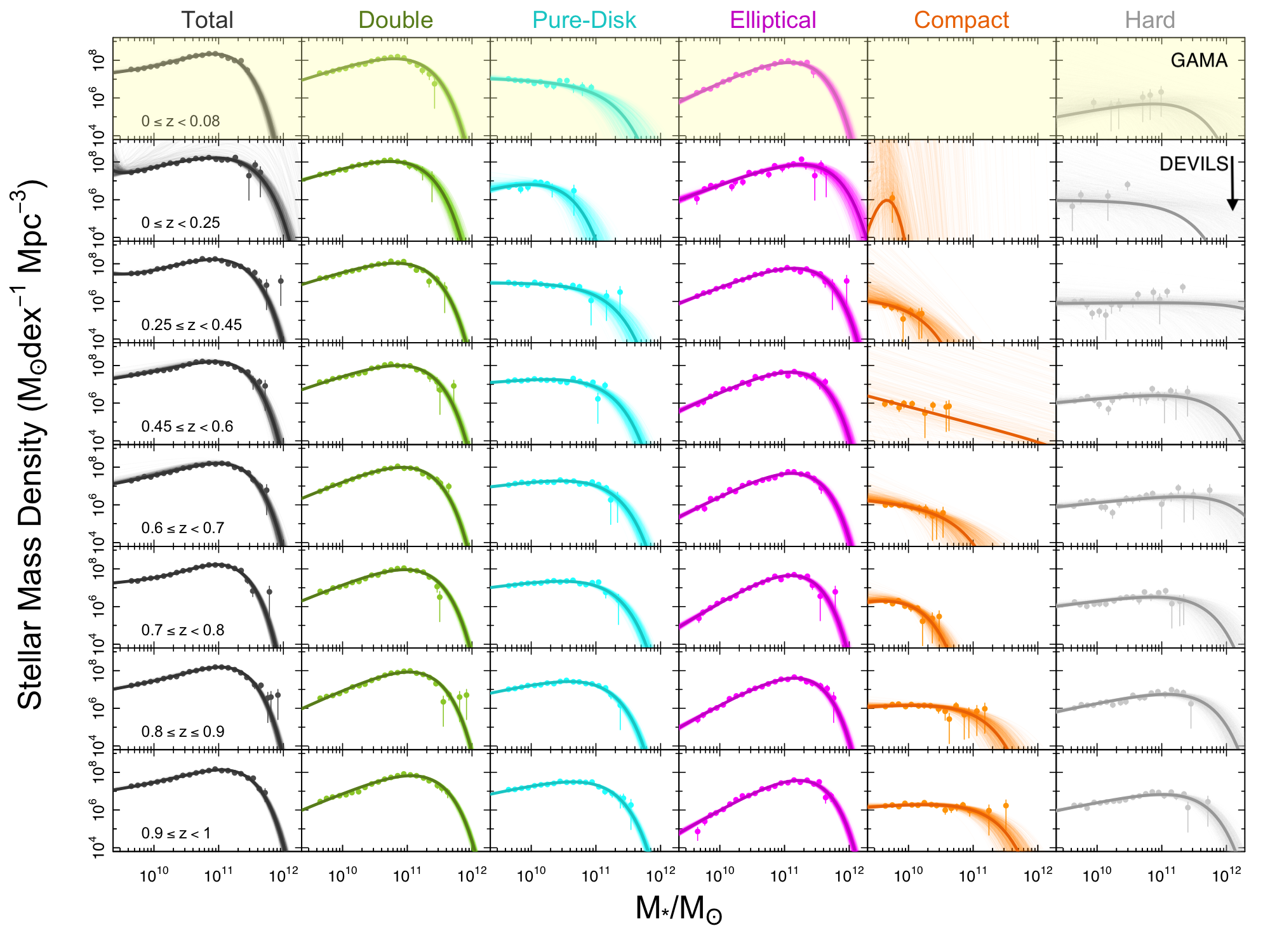}
	\caption{The distribution of \textit{total} and morphological stellar mass density in different redshifts. Redshift bins are the same as Figure \ref{fig:Mfunc_6z}. Points and lines indicate $M^\prime \Phi(M^\prime)$ in Equation \ref{eq:massDens}, where $\Phi(M^\prime)$ is our Schechter function fit. The shaded transparent regions represent the error range calculated by 1000 times sampling of the full posterior probability distribution of the fit parameters. The distribution of the stellar mass density of all individual morphological types in almost all redshift bins is bounded. Top row highlighted by yellow colour represents the GAMA data ($0 \le z \le 0.08$).}
	\label{fig:Mdens_6z}
 \end{figure}
\end{landscape}

In this section, we investigate the evolution of the Stellar Mass Density (SMD) as a function of morphological types. To determine the SMD, we integrate under the best fit Schechter functions over all stellar masses from $10^{9.5}$ to $\infty$. This integral can be expressed as a gamma function:

\begin{equation}
\rho_* = \int_{M=10^{9.5}}^{\infty} M^\prime \Phi(M^\prime) dM^\prime = \Phi^* M^* \Gamma(\alpha+2,10^{9.5}/M^*),
\label{eq:massDens}
\end{equation}

\noindent where $\Phi^*$, $M^*$ and $\alpha$ are the best regressed Schechter parameters. Figure \ref{fig:Mdens_6z} shows the evolution of the distribution of the SMDs, term $M^\prime \Phi(M^\prime)$, for different morphologies. We also illustrate the fitting errors by sampling 1000 times the full posterior probability distribution of the fit parameters. These are shown in Figure \ref{fig:Mdens_6z} as transparent shaded regions around the best fit curves. The standard deviations of the integrated SMD calculated from each of these functions are reported as the fit error on $\rho_*$ in Table \ref{tab:rho}. As can be seen in Figure \ref{fig:Mdens_6z}, the distribution of the stellar mass density of all individual morphological types in almost all redshift bins is bounded, implying that integrating under these curves will capture the majority of the stellar mass for each class. 

\begin{figure*}
	\centering
	\includegraphics[width = \textwidth, angle = 0]{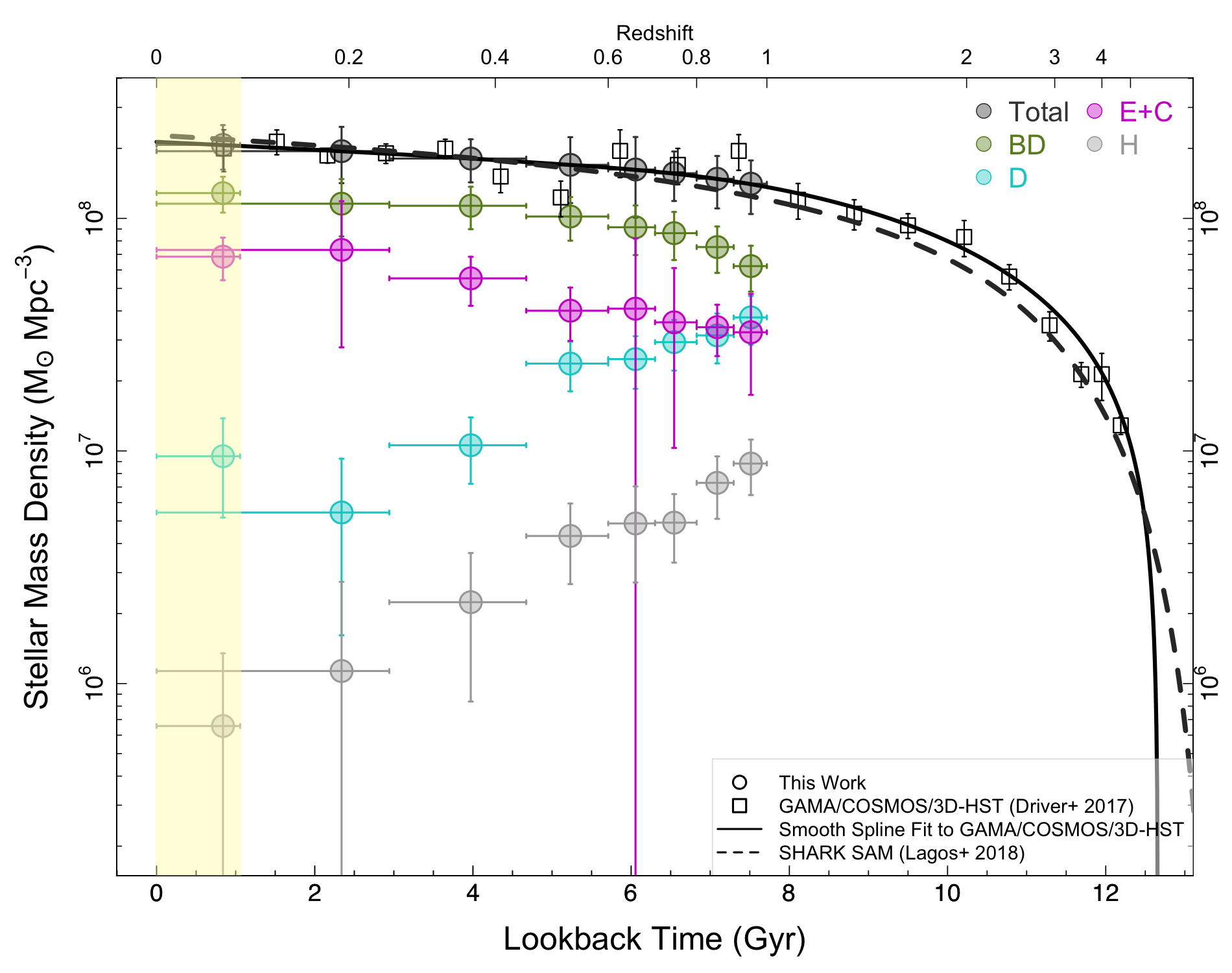}
	\caption{The evolution of the \textit{total} and morphological stellar mass density, $\rho_{*}$, in the last 8 Gyr of the cosmic age. $\rho_{*}$ expresses our measurements of the analytical integration under the best Schechter function fits in 8 redshift bins, i.e. Equation \ref{eq:massDens}. Highlighted region shows the epoch covered by the GAMA data ($0 \le z \le 0.08$). This figure includes the correction for the large scale structure, LSS, that we apply by fitting a smooth spline to the empirical data of GAMA/G10COSMOS/3D-HST by \protect\cite{Driver18}. See the text for more details on our method for this correction. The prediction of the SHARK semi-analytic model is also overlaid as dashed line. Colour codes are similar to Figure \ref{fig:Mfunc_6z}. Vertical bars on the points show the total error budget on each data point including: cosmic variance within the associated redshift bin taken from \protect\cite{Driver10}, classification error, fit error and Poisson error. See Section \ref{sebsec:morph_err_budget} for more details about our error analysis. Horizontal bars indicate the redshift ranges while the data points are plotted at the mean redshift.}
	\label{fig:MassBuildUp_LSS}
\end{figure*}

Figure \ref{fig:MassBuildUp_LSS} then shows the evolution of the integrated SMD, $\rho_*$, in the Universe between $z = 0-1$. 
This includes the LSS correction by forcing our \textit{total} SMD values to match the smooth spline fit to the \cite{Driver18} data, as described in Section \ref{subsec:LSS_cor}. The uncorrected evolutionary path of $\rho_*$ is shown in the upper panel of Figure \ref{fig:LSS_cor}.

We report the empirical LSS corrected $\rho_*$ values in Table \ref{tab:rho}. This table also provides the LSS correction factor, $\beta$, so one might obtain the original values by $\rho_*^{Orig} = \rho_*^{corr}/\beta$. 

For completeness, we show the evolution of the SMDs before we apply the LSS corrections in Figure \ref{fig:SMD_evol_noLSS} of Appendix \ref{sec:SMD_evol_noLSS}. In this Figure, the trends are not as smooth as one would expect without the LSS correction being applied, given the evident structure in the $N(z)$ distribution from Figure \ref{fig:zdens}. However, even without the LSS corrections, the main trends are still present, albeit not as strong. This highlights that the LSS correction is an important aspect and also the need for much wider deep coverage than that currently provided by HST. This will become possible in the coming Euclid and Roman era.

Before we analyse the evolution of the SMDs in Figure \ref{fig:MassBuildUp_LSS}, below we investigate the errors that are involved in this calculation.

\subsection{Analysing the Error Budget on the SMD}
\label{sebsec:morph_err_budget}

The error budget on our analysis of the SMD includes \textit{cosmic variance } (CV), \textit{fit error}, \textit{classification error} and \textit{Poisson error}. Cosmic variance and classification are the dominant sources of error. 

We make use of \cite{Driver10} equation 3 to calculate the \textit{cosmic variance} in the volume encompassed within each redshift bin. This calculation is implemented in the R package: {\fontfamily{qcr}\selectfont celestial}\footnote{ The online version of this cosmology calculator is available at: \href{http://cosmocalc.icrar.org/}{http://cosmocalc.icrar.org/} }. Note that for our low-$z$ GAMA data, instead of using \cite{Driver10} equation that estimates a $\sim 22.8\%$ cosmic variance, we empirically calculate the CV using the variation of the SMD between 3 different GAMA regions of G09, G12, and G15 with a total effective area of 169.3 square degrees (G09:54.93; G12: 57.44; G15: 56.93, \citealt{Bellstedt20a}) and find CV to be $\sim 16\%$. 

We measure the \textit{fit error} by 1000 times sampling of the full posterior probability distribution of the Schechter parameters (shown as shaded error regions in Figure \ref{fig:Mdens_6z}) and calculating the associated $\rho_*$ in each iteration. 
We calculate the \textit{classification error} by measuring the stellar mass density associated with each of our 3 independent morphological classifications. The range of variation of the SMD between classifiers gives the error of our classification. 
The \textit{Poisson error} is calculated by using the number of objects in each morphology per redshift bin.

The combination of all the above error sources will provide us with the total error that is reported in Table \ref{tab:rho}.
\\[2\baselineskip]


As can be seen in Figure \ref{fig:MassBuildUp_LSS}, the extrapolation of our D10/ACS $\rho_*$ to $z = 0$ agrees well with the our local GAMA estimations. Note slight difference in D, but consistent in errors.

\begin{figure}
	\centering
	\includegraphics[width=0.49\textwidth]{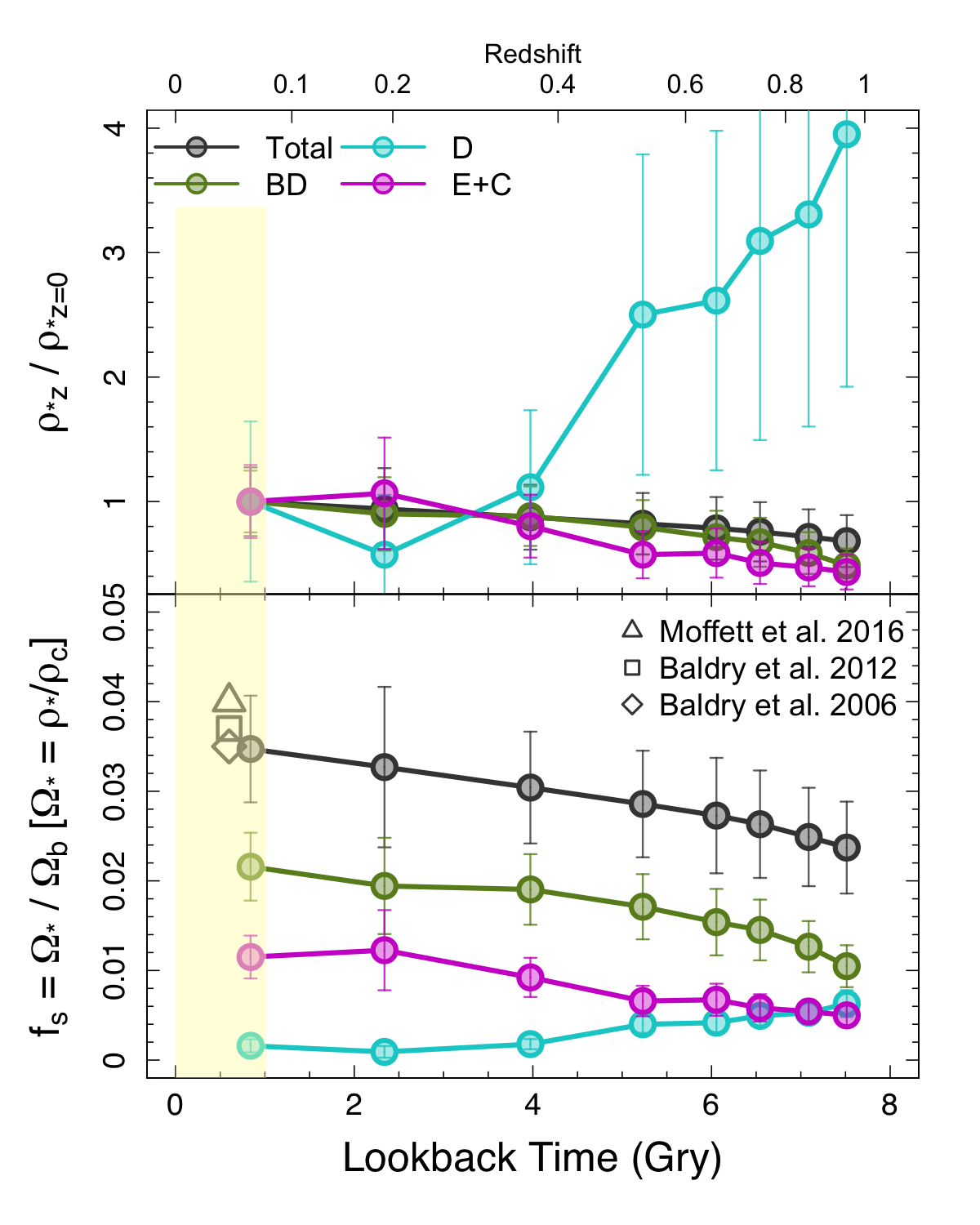}
	\caption{Top panel: the variation of the stellar mass density showing the fraction of final stellar mass density assembled or lost by each redshift, i.e. $\rho_{*z}/\rho_{*z=0}$. $\rho_{*z=0}$ is taken from GAMA estimations of the local Universe. Bottom panel: the evolution of the baryon fraction in form of stars ($f_s$). For simplicity, in this figure we only show the main morphological types and remove the \textit{hard} and \textit{compact} sub-classes. Highlighted region shows the epoch covered by the GAMA data.}
	\label{fig:MassDensVar_fs}
\end{figure}

\begin{figure*}
\centering
\begin{subfigure}
  \centering
  \includegraphics[width=0.49\textwidth]{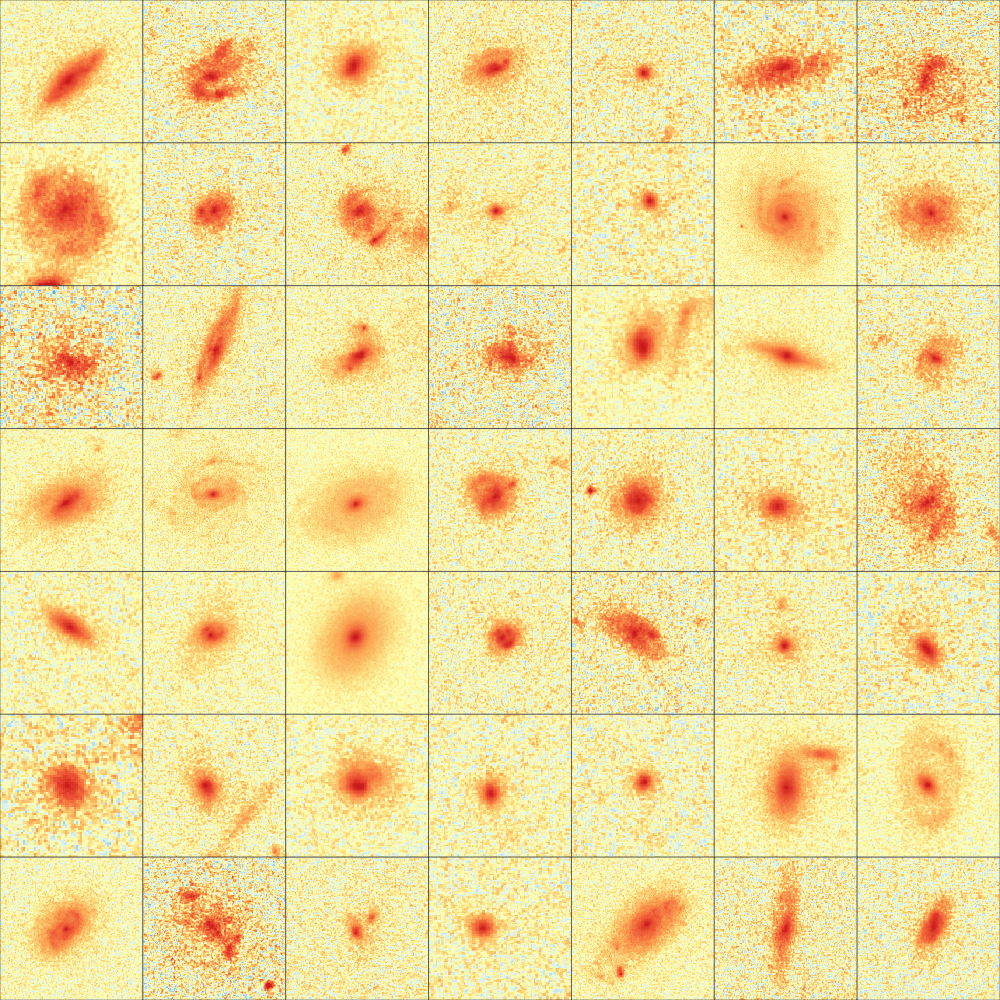}
  \label{subfig:DoubleStamps_F814W}
\end{subfigure}%
\begin{subfigure}
  \centering
  \includegraphics[width=0.49\textwidth]{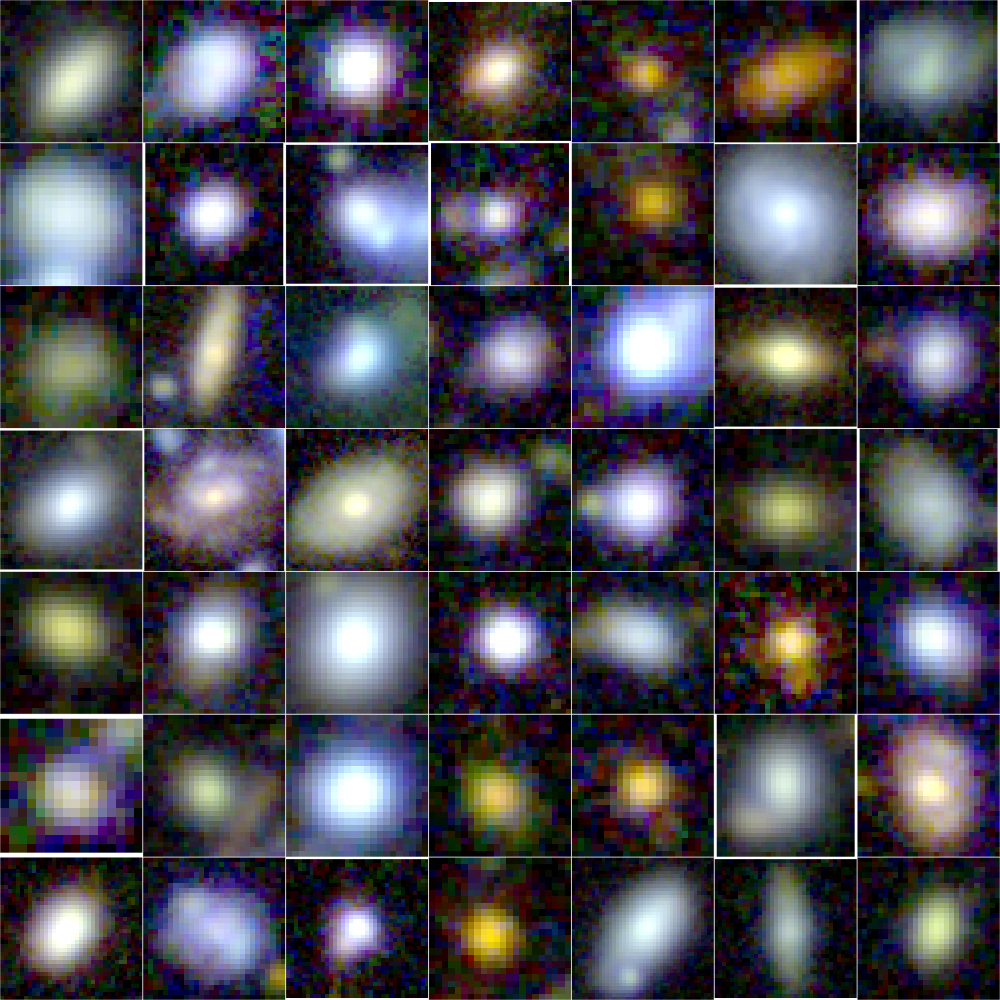}
  \label{subfig:DoubleStamps_RGB}
\end{subfigure}
\caption{Random sample of double component systems. Left set of panels: ACS/F814W image. Right set of panels: SUBARU $gri$ combined image.}
\label{fig:contact_sheets_double}
\end{figure*}

\begin{figure*}
\centering
\begin{subfigure}
  \centering
  \includegraphics[width=0.49\textwidth]{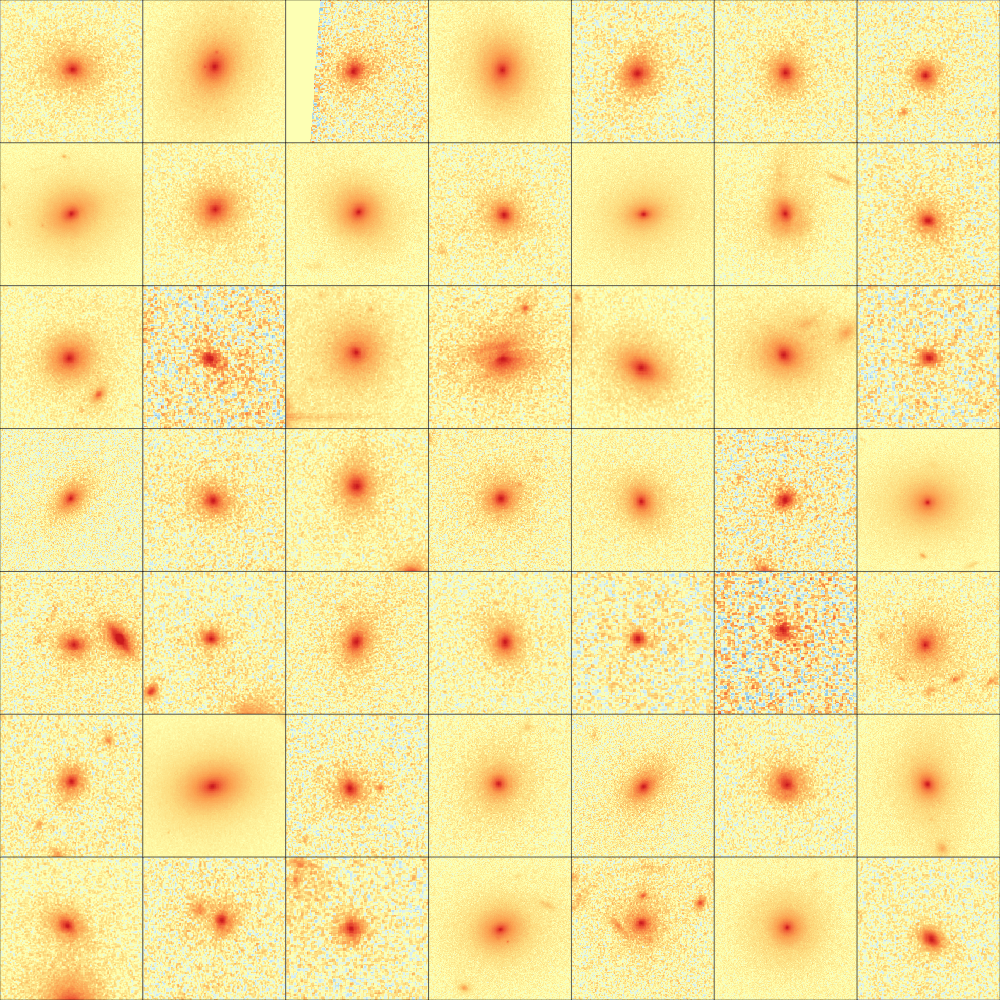}
  \label{subfig:EllipStamps_F814W}
\end{subfigure}%
\begin{subfigure}
  \centering
  \includegraphics[width=0.49\textwidth]{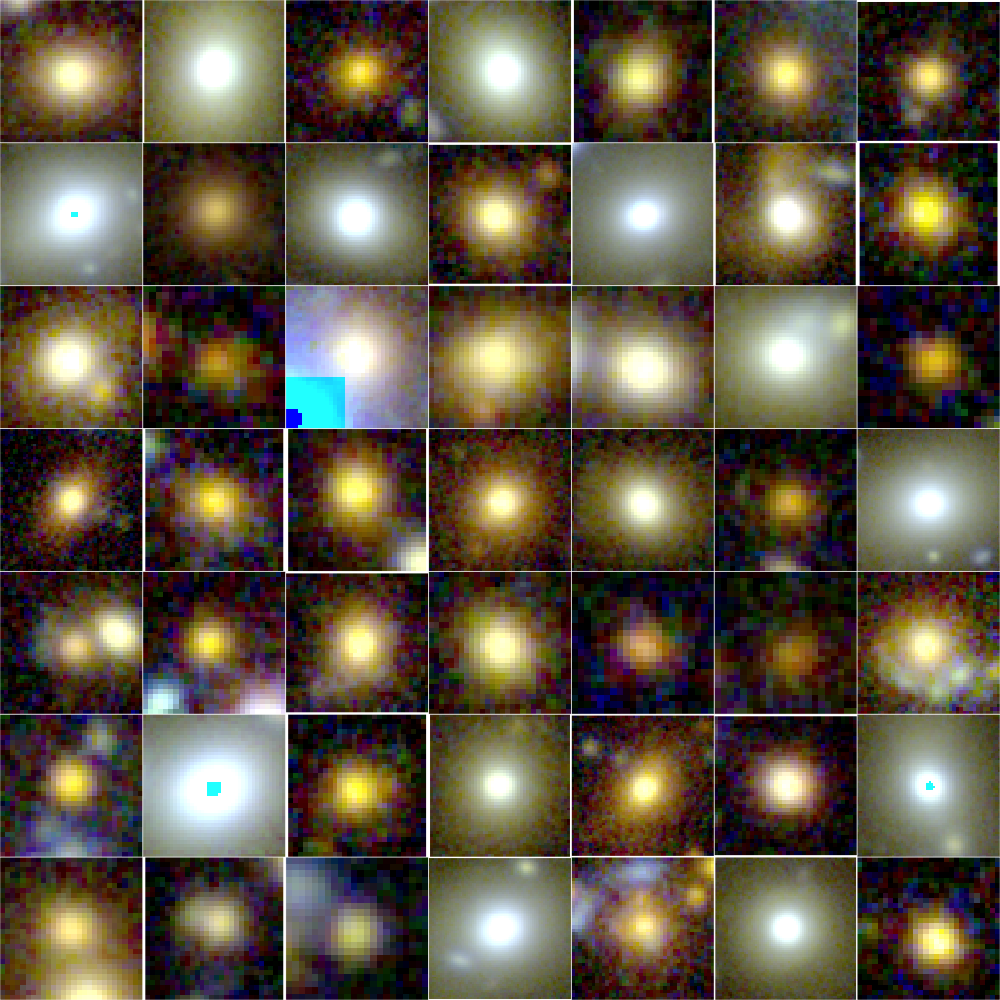}
  \label{subfig:EllipStamps_RGB}
\end{subfigure}
\caption{Random sample of elliptical systems. Left set of panels: ACS/F814W image. Right set of panels: SUBARU $gri$ combined image.}
\label{fig:contact_sheets_ellip}
\end{figure*}

\begin{figure*}
\centering
\begin{subfigure}
  \centering
  \includegraphics[width=0.49\textwidth]{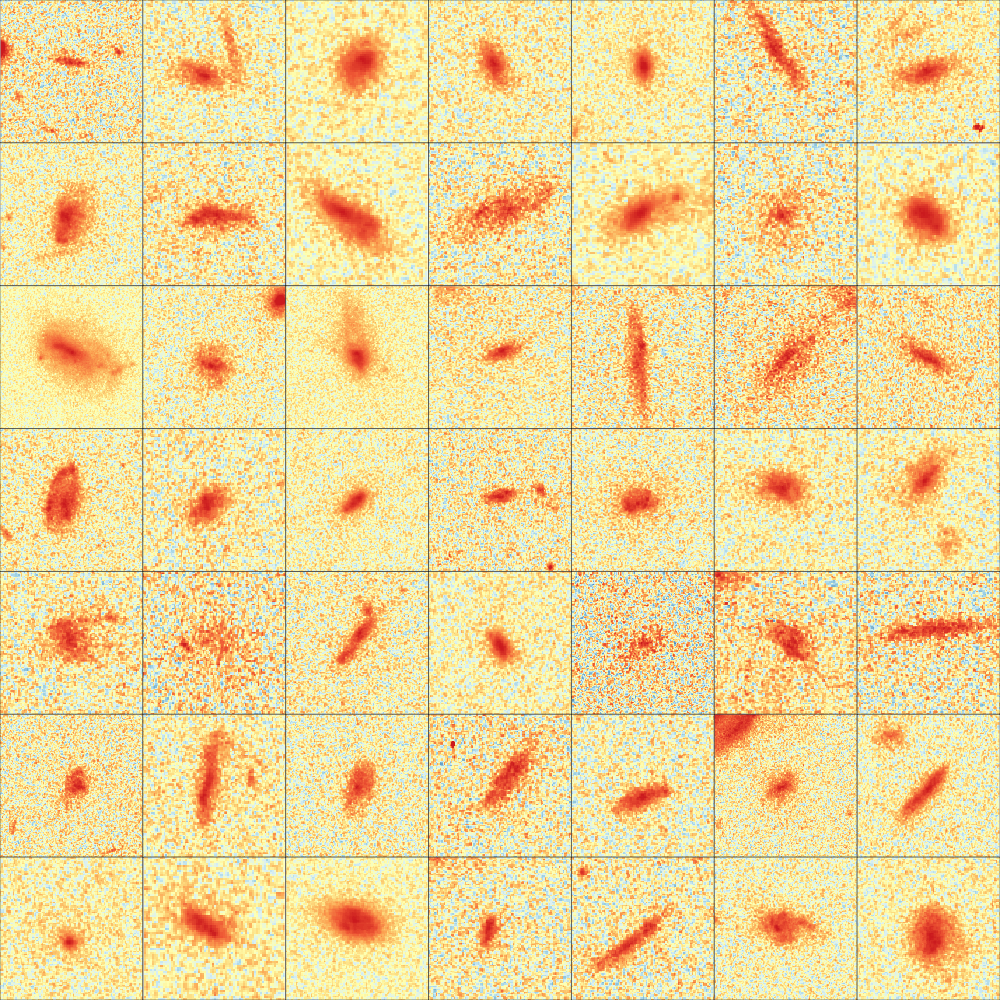}
  \label{subfig:DiskStamps_F814W}
\end{subfigure}%
\begin{subfigure}
  \centering
  \includegraphics[width=0.49\textwidth]{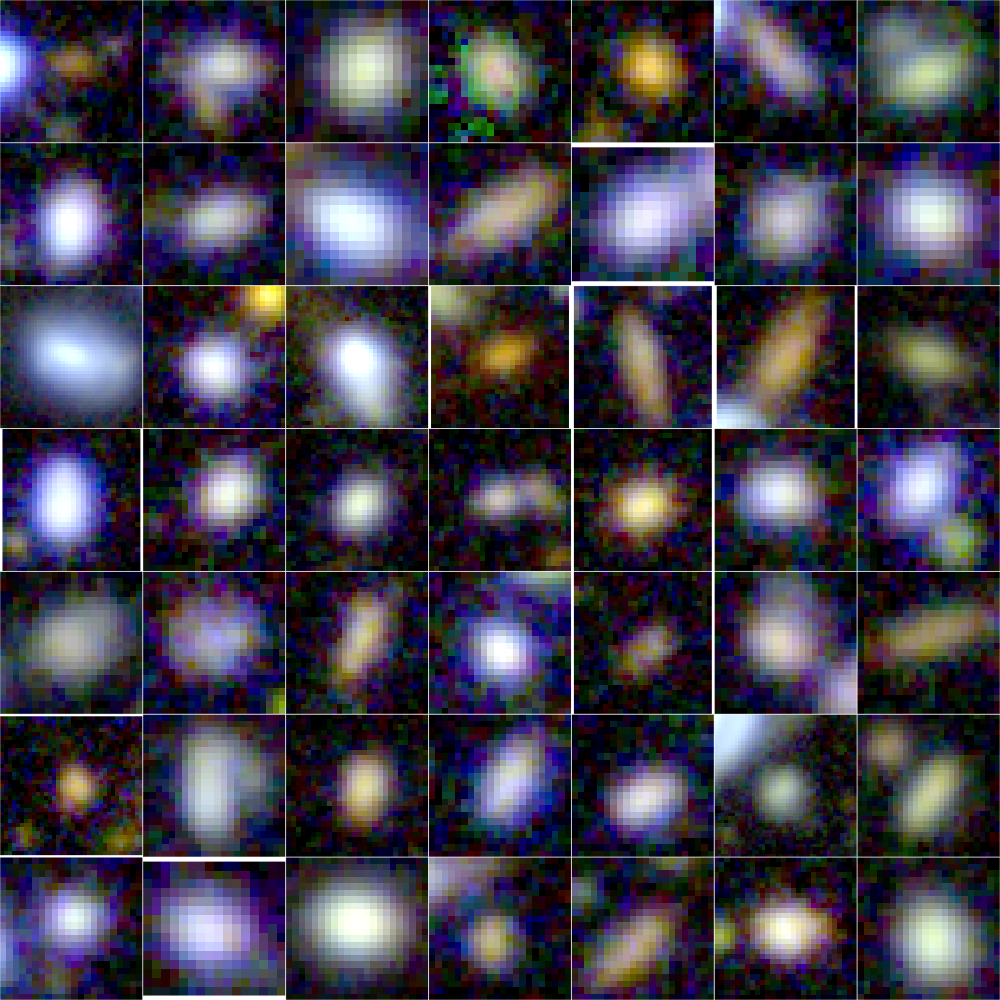}
  \label{subfig:DiskStamps_RGB}
\end{subfigure}
\caption{Random sample of pure disk systems. Left set of panels: ACS/F814W image. Right set of panels: SUBARU $gri$ combined image.}
\label{fig:contact_sheets_disk}
\end{figure*}

\begin{figure*}
\centering
\begin{subfigure}
  \centering
  \includegraphics[width=0.49\textwidth]{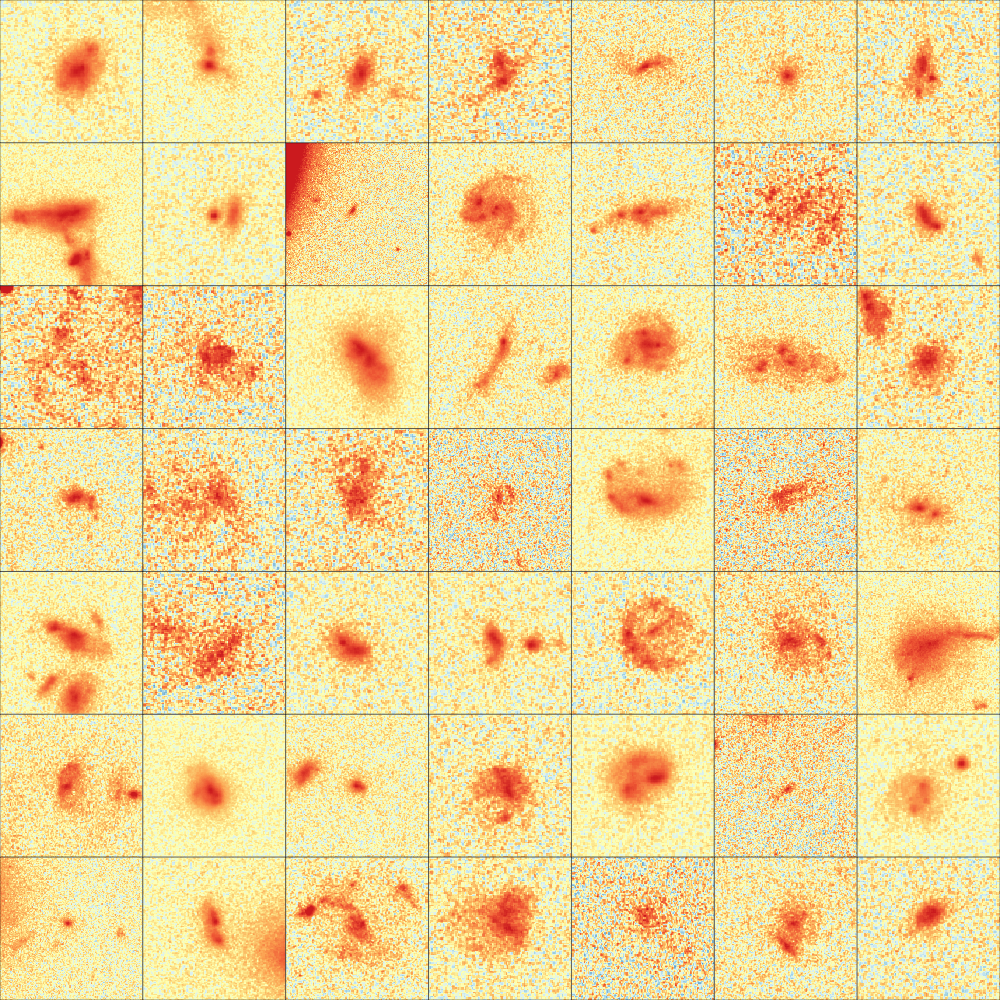}
  \label{subfig:HardStamps_F814W}
\end{subfigure}%
\begin{subfigure}
  \centering
  \includegraphics[width=0.49\textwidth]{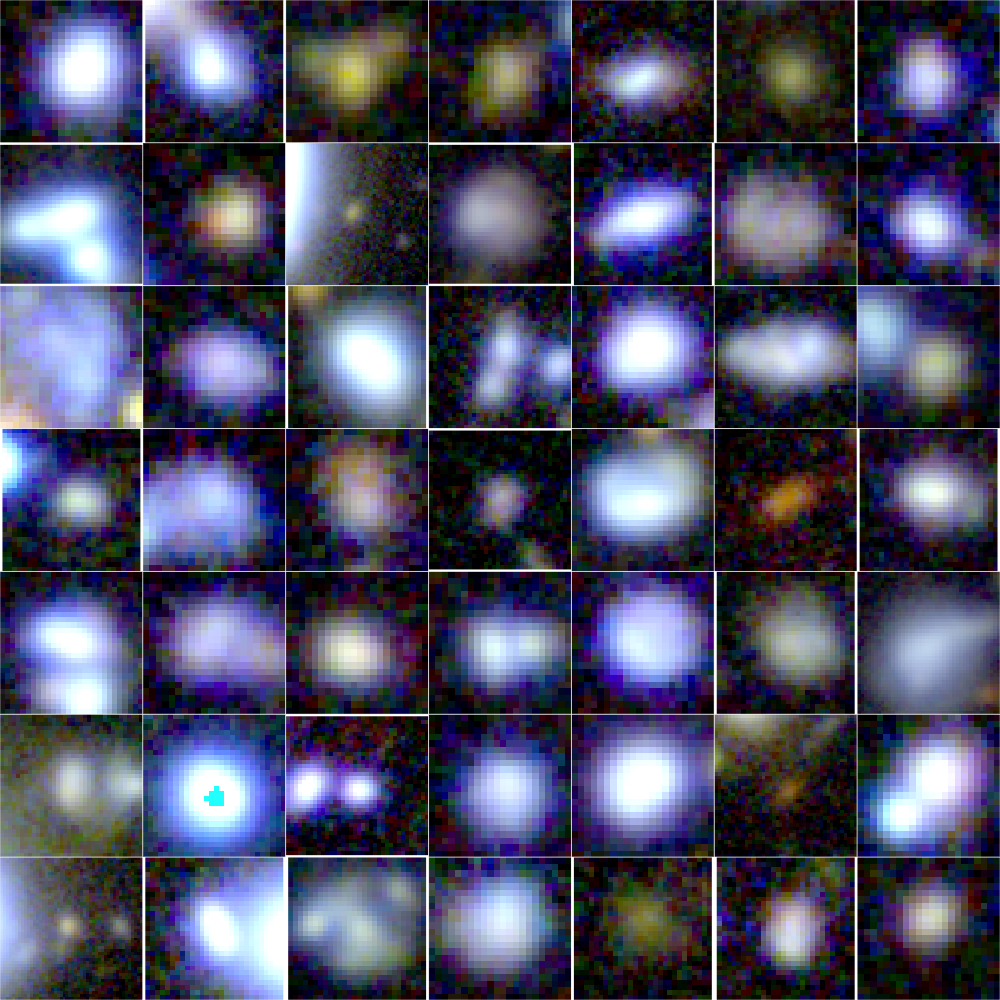}
  \label{subfig:HardStamps_RGB}
\end{subfigure}
\caption{Random sample of complex systems (\textit{hard}). Left set of panels: ACS/F814W image. Right set of panels: SUBARU $gri$ combined image.}
\label{fig:contact_sheets_hard}
\end{figure*}

\begin{figure*}
\centering
\begin{subfigure}
  \centering
  \includegraphics[width=0.49\textwidth]{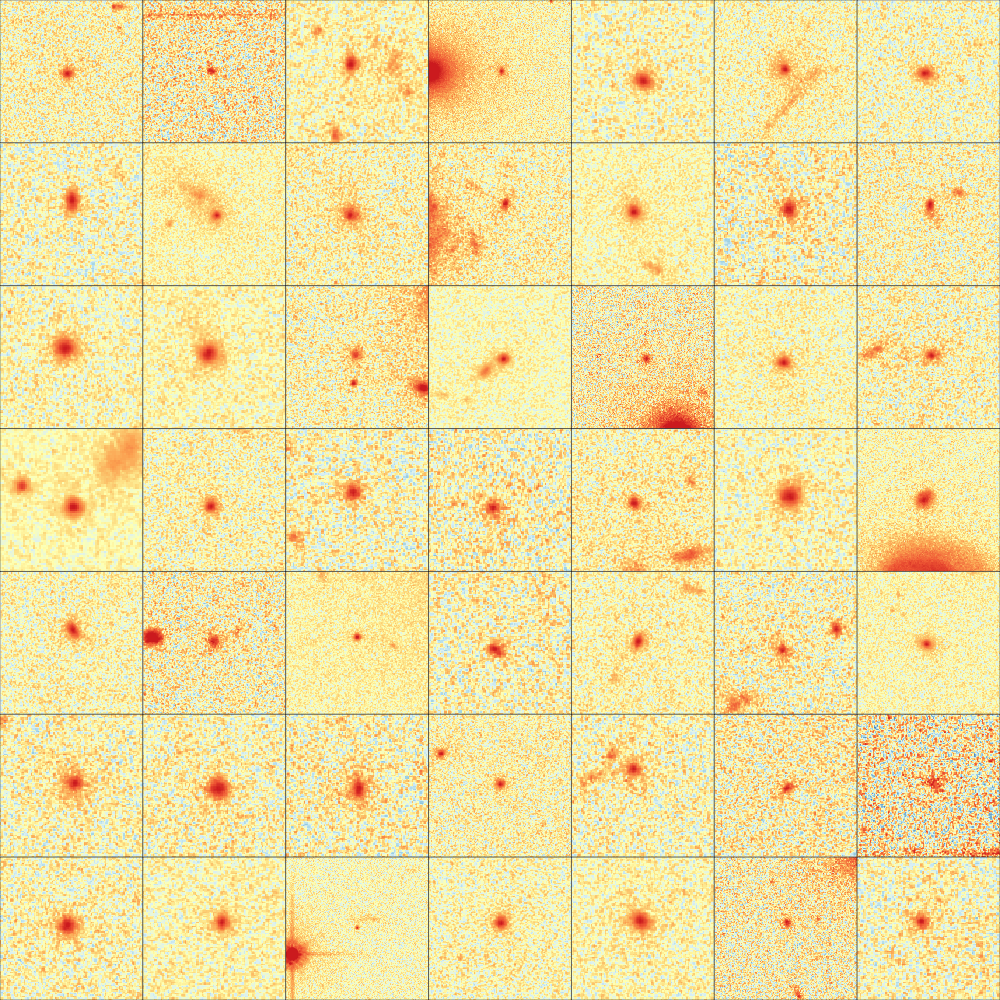}
  \label{subfig:CompStamps_F814W}
\end{subfigure}%
\begin{subfigure}
  \centering
  \includegraphics[width=0.49\textwidth]{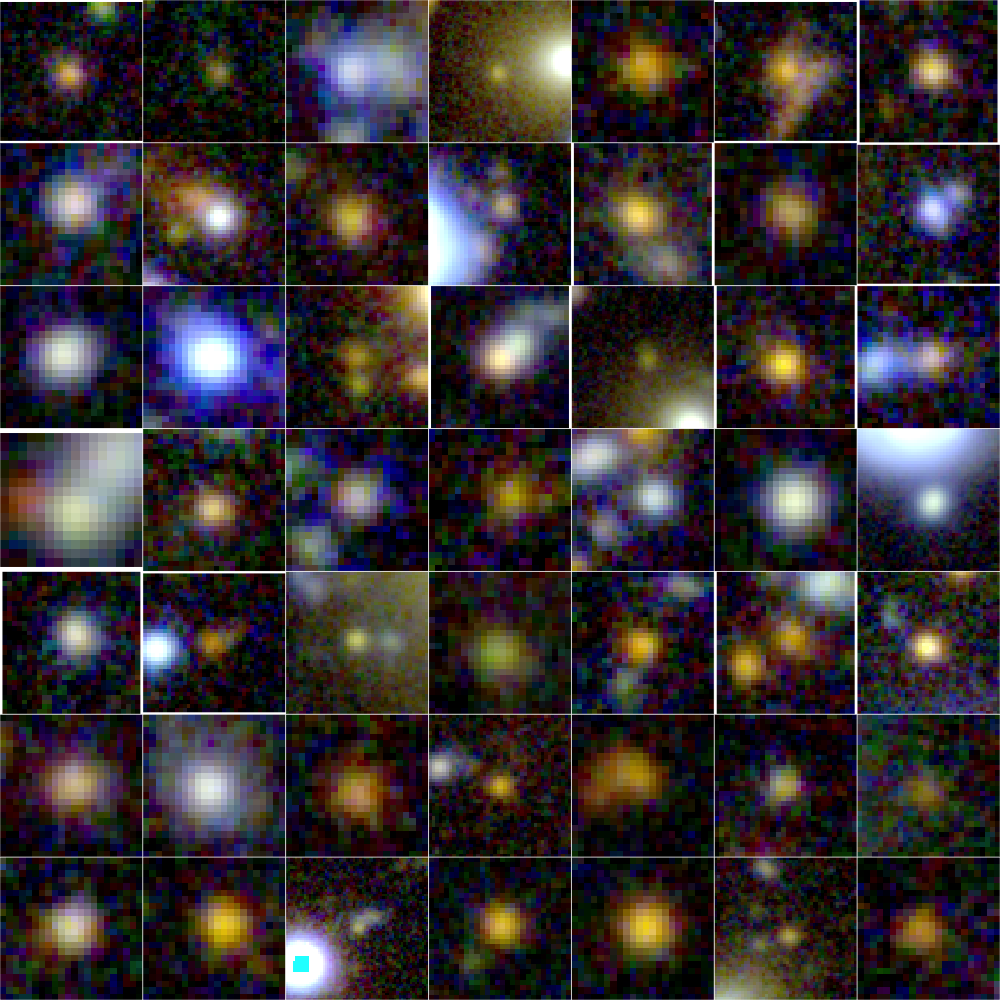}
  \label{subfig:CompStamps_RGB}
\end{subfigure}
\caption{Random sample of low angular sized systems (\textit{compact}). Left set of panels: ACS/F814W image. Right set of panels: SUBARU $gri$ combined image.}
\label{fig:contact_sheets_comp}
\end{figure*}

The total change in the stellar mass is consistent with observed SFR evolution (e.g., \citealt{Madau14}; \citealt{Driver18}) as we will discuss more in Section \ref{sec:discussion}. Analysing the evolution of the SMD (Figure \ref{fig:MassBuildUp_LSS}), we find that in total (black symbols), 68\% of the current stellar mass in galaxies was in place $\sim 8$ Gyr ago ($z \sim 1.0$). The top panel of Figure \ref{fig:MassDensVar_fs} illustrates the variation of the $\rho_*$ ($\rho_{z}^{*} / \rho_{z=0}^{*}$), where $\rho_{z}^{*}$ is the SMD at redshift $z$ while $\rho_{z=0}^{*}$ represents the final SMD at $z~=~0$. 

According to our visual inspections C types are closer to Es than other subcategories. We, therefore, combine C types with E galaxies that shows a smooth growth with time of a factor of $\sim 2.5$ up to $z = 0.25$ and flattens out since then ($0.0 < z < 0.25$). Note that as reported in Table \ref{tab:rho} the amount of mass in the C class is very little. This demonstrates a significant mass build-up in E galaxies over this epoch ($\sim 150\%$). We also note that the large error bars on some of E+C data points are dominantly driven by the large errors in C SMDs.  

Having analytically integrated the SMD, we now measure the fraction of baryons in the form of stars ($f_s = \Omega_*/\Omega_b$) locked in each of our morphological types. We adopt $\Omega_b = 0.0493$ as estimated from Planck by \cite{Planck20} and the critical density of the Universe at the median redshift of GAMA, i.e. $z = 0.06$ to be $\rho_c = 1.21 \times 10^{11} \mathrm{M}_{\odot} \mathrm{Mpc}^{-3}$ in a 737 cosmology.

As shown in the bottom panel of Figure \ref{fig:MassDensVar_fs}, at our D10/ACS lowest redshift bin $\overline{z} \sim 0.18$ we find the fraction of baryons in stars $f_s \sim 0.033\pm0.009$. Including our GAMA measurements at $\overline{z} \sim 0.06$ we find this fraction to be $f_s \sim 0.035\pm0.006$. As shown in Figure \ref{fig:MassDensVar_fs}, this result is consistent with other studies within the quoted errors for example: \cite{Baldry06}, \cite{Baldry12} and \cite{Moffett16a}.
The evolution of the fraction of baryons locked in stars, $f_s$, shows that it has increased from $(2.4\pm0.5)$\% at $z \sim 1$ to $(3.5\pm0.6)$\% at $z \sim 0$ indicating an increase of a factor of $\sim 1.5$ during last 8 Gyr.   

Figure \ref{fig:MassDensVar_fs} also shows the breakdown of the total $f_s$ to each of our morphological subcategories highlighting that as expected BD systems contribute the most to the stellar baryon fraction increasing from $f_s = 0.011\pm0.006$ to $f_s = 0.022\pm0.005$ at $0 < z < 1$.
E+C systems take less percentage of the total $f_s$ but increase their contribution from $0.005\pm0.001$ to $0.012\pm0.002$ while D galaxies decrease from $f_s = 0.006\pm0.002$ to $f_s = 0.001\pm0.005$ between $0 < z < 1$.
We report our full $f_s$ values for all morphological types at all redshifts in Table. \ref{tab:fs}.

In summary, over the last 8 Gyr, double component galaxies clearly dominate the overall stellar mass density of the Universe at all epochs. The second dominant system is E galaxies. However the extrapolation of the trends to higher redshifts in Figures \ref{fig:MassBuildUp_LSS} and \ref{fig:MassDensVar_fs} indicates that D systems are likely to dominate over Es in the very high-$z$ regime ($z > 1$), which is reasonable according to the rise of the cosmic star formation history and the association of star-formation with disks. 

\begin{table*}
\centering
\caption{The \textit{total} and morphological stellar mass density in different redshift ranges displayed in Figure \ref{fig:MassBuildUp_LSS} (left panel). $\rho_*$ values are calculated from the integration under the best Schechter function fits. Note that the $\rho_*$ are presented after applying the LSS correction. Columns represent: $z-$bin: redshift bins, $\overline{z}$: mean redshift, LBT: lookback time, T (S-Schechter): \textit{total with single Schechter}, T (D-Schechter): \textit{total with double Schechter}, BD: \textit{bulge+disk}, pD: \textit{pure-disk}, E: \textit{elliptical}, C: \textit{compact}, H: \textit{hard}, $\beta$: the large scale structure correction factor. Errors incorporate all error sources including, cosmic variance, fit error, Poisson error and classification error (see Section \ref{sebsec:morph_err_budget} for details). }
\begin{adjustbox}{angle = 0, scale = 0.8}
\begin{tabular}{ccccccccccccc}
\firsthline \firsthline \\
                      &     &   &  \multicolumn{6}{c}{log10$(\rho_*/M_\odot) $} \\ \\
\cline{4-9} \\
$z-$bin      & $\overline{z}$ & LBT (Gyr) & T (S-Schechter)  & T (D-Schechter)  & BD               & pD               & E                & C                & H                & $\beta$ \\ \\ \hline \\
$0.00 \le z < 0.08$   &  $0.06$ & $0.82$  &  $8.315\pm0.08$  &  $8.316\pm0.11$  &  $8.109\pm0.08$  &  $6.977\pm0.26$  &  $7.835\pm0.10$  &  $-$             &  $5.818\pm1.25$  &  $1.40$        \\
$0.00 \le z < 0.25$   &  $0.18$ & $2.27$  &  $8.289\pm0.14$  &  $8.289\pm0.14$  &  $8.063\pm0.14$  &  $6.735\pm0.53$  &  $7.863\pm0.20$  &  $5.332\pm0.83$  &  $6.054\pm0.38$  &  $0.89$        \\
$0.25 \le z < 0.45$   &  $0.36$ & $3.95$  &  $8.258\pm0.10$  &  $8.257\pm0.10$  &  $8.054\pm0.10$  &  $7.025\pm0.17$  &  $7.740\pm0.12$  &  $5.513\pm1.23$  &  $6.350\pm0.43$  &  $0.74$        \\
$0.45 \le z < 0.60$   &  $0.53$ & $5.23$  &  $8.231\pm0.10$  &  $8.230\pm0.16$  &  $8.008\pm0.10$  &  $7.376\pm0.12$  &  $7.594\pm0.13$  &  $5.936\pm1.01$  &  $6.635\pm0.21$  &  $1.21$       \\
$0.60 \le z < 0.70$   &  $0.65$ & $6.05$  &  $8.210\pm0.12$  &  $8.210\pm0.21$  &  $7.962\pm0.12$  &  $7.395\pm0.13$  &  $7.603\pm0.13$  &  $5.922\pm0.60$  &  $6.688\pm0.25$  &  $0.74$       \\
$0.70 \le z < 0.80$   &  $0.74$ & $6.55$  &  $8.195\pm0.11$  &  $8.194\pm0.12$  &  $7.936\pm0.12$  &  $7.468\pm0.12$  &  $7.540\pm0.13$  &  $6.032\pm0.95$  &  $6.692\pm0.17$  &  $0.81$       \\
$0.80 \le z < 0.90$   &  $0.85$ & $7.07$  &  $8.171\pm0.11$  &  $8.170\pm0.13$  &  $7.877\pm0.11$  &  $7.497\pm0.12$  &  $7.508\pm0.12$  &  $6.258\pm0.55$  &  $6.863\pm0.15$  &  $0.57$       \\
$0.90 \le z \le 1.00$ &  $0.95$ & $7.51$  &  $8.150\pm0.11$  &  $8.149\pm0.13$  &  $7.795\pm0.11$  &  $7.574\pm0.12$  &  $7.473\pm0.12$  &  $6.430\pm0.43$  &  $6.946\pm0.14$  &  $0.55$       \\
\\

\lasthline

\end{tabular}
\end{adjustbox}
\label{tab:rho}
\end{table*}

\section{Discussion} 
\label{sec:discussion}

Making use of our morphological classifications, we explore the evolution of the stellar mass function at $0 \le z \le 1$, to assess the physical processes that are likely affecting the individual morphological SMF. In particular, major mergers are thought to primarily occur between comparable mass companions (1:3) resulting in the fast growth of the SMF \citep{Robotham14}. Conversely, secular activities, minor mergers and tidal interactions will primarily alter the number density at the low-stellar mass end of the SMF \citep{Robotham14}. Investigating the \textit{total} SMF (Figure \ref{fig:Mfunc_6z}) shows that unlike high-mass end the low-mass end grows significantly. This suggests that since $z~\sim~1$, the galaxy population goes through mainly in-situ/secular processes at the low-mass end.

An important caveat, in what follows, is that although we have undertaken multiple tests of our visual classifications, possible uncertainties due to human classification error or inconsistencies will be present. Nevertheless some clear trends, which we believe are resilient to the classifications uncertainties are evident. Furthermore, we do not rule out the effects of dust in distinguishing bulges, particularly at high-$z$ leading to overestimating the number of \textit{pure-disk} systems at high redshifts. Although, high level of agreement in our visual classifications (see Figure \ref{fig:agreent}) gives us some confidence that our evolutionary trends are unlikely dominated by incorrect classifications, it is possible that all classifiers agree on the incorrect classification.

Firstly, double component galaxies display a modest decrease with redshift in the high-stellar mass end of their SMF (Figure \ref{fig:Mfunc_6z}), whilst their low-stellar mass end steepens significantly. This could be interpreted as most of the stellar mass in double component systems evolving via lower mass weighted star formation or secular activity, rather than major merging, i.e., BD systems are not merging together to form higher-mass BD systems. 

Secondly, \textit{pure-disk} systems show a strong increase at the low-mass end and a {\it decrease} at their high-mass end. The low-mass end evolution suggests in-situ star formation of the disk and/or the formation of new disks. However, the decrease at the high-mass end is to some extent unphysical, unless these systems are undergoing a transformation from the D class to another class. The most likely prospect is the secular formation of a central bulge component, resulting in a morphological transformation into the BD class. Hence, as time progresses and the second component forms, galaxies exit the D class leading to a mass-deficit at the high-mass end.

The new component forming through such an in-situ process is most likely a pseudo-bulge (pB), resulting in mass loss from the high-mass D SMF and a corresponding mass gain at comparable mass in the BD class. 

Finally, \textit{elliptical} galaxies, generally thought to be inert, show little growth in their high-mass end but a significant growth of low- and intermediate mass end presumably due to mergers.  

The evolution of the global and morphological SMDs indicates that BD galaxies dominate ($\sim 60\%$ on average) the stellar mass density of the Universe since at least at $z < 1$. This morphological class also shows a constant mass growth, compared to other morphologies. As mentioned, D systems slightly decrease their stellar mass density with time. Presumably, the mass transfer out of the D class out-weighs the mass gain due to in-situ star-formation for this class. 

We remark that the \textit{rate} of mass growth in the BD systems also decreases with time. This is reflected in the decrease of their SMD slope although it is still steeper than the \textit{total} SMD evolution and consistent with the general decline in the cosmic star-formation history. On the other hand, the E galaxies experience an initial growth until $z\sim0.2$ and a recent flattening in their stellar-mass growth from $z=1-0$ (see Figure \ref{fig:MassBuildUp_LSS} and the top panel of Figure \ref{fig:MassDensVar_fs}). 

\begin{table*}
\centering
\caption{\textit{Total} and morphological stellar baryon fraction ($f_s$) in different times. See the text for details.}
\begin{adjustbox}{angle = 0, scale = 0.8}
\begin{tabular}{lcccccccc}
\firsthline \firsthline \\
                      &   \multicolumn{7}{c}{Redshift}  \\ \\
\cline{2-8} \\

Morphology Type         & $0.00 \le z < 0.08$    & $0.00 \le z < 0.25$ & $0.25 \le z < 0.45$ & $0.45 \le z < 0.60$ & $0.60 \le z < 0.70$ & $0.70 \le z < 0.80$ & $0.80 \le z < 0.90$ & $0.90 \le z < 1.0$ \\ \hline \\
Total                & $0.035\pm0.006$ & $0.033\pm0.009$ & $0.030\pm0.006$ & $0.029\pm0.006$ & $0.027\pm0.006$ & $0.026\pm0.006$ & $0.025\pm0.005$ & $0.024\pm0.005$ \\
Double               & $0.022\pm0.004$ & $0.019\pm0.005$ & $0.019\pm0.004$ & $0.017\pm0.004$ & $0.015\pm0.004$ & $0.015\pm0.003$ & $0.013\pm0.003$ & $0.010\pm0.002$ \\
Pure-Disk            & $0.002\pm0.001$ & $0.001\pm0.001$ & $0.002\pm0.001$ & $0.004\pm0.001$ & $0.004\pm0.001$ & $0.005\pm0.001$ & $0.005\pm0.001$ & $0.006\pm0.001$ \\
Elliptical           & $0.012\pm0.002$ & $0.012\pm0.004$ & $0.009\pm0.002$ & $0.007\pm0.002$ & $0.007\pm0.002$ & $0.006\pm0.002$ & $0.005\pm0.001$ & $0.005\pm0.001$ \\
Compact              & $-$             & $0.000\pm0.000$ & $0.000\pm0.000$ & $0.000\pm0.000$ & $0.000\pm0.000$ & $0.000\pm0.000$ & $0.000\pm0.000$ & $0.000\pm0.000$ \\
E+C                  & $0.012\pm0.002$ & $0.012\pm0.004$ & $0.009\pm0.002$ & $0.007\pm0.002$ & $0.007\pm0.002$ & $0.006\pm0.002$ & $0.006\pm0.001$ & $0.005\pm0.001$ \\
Hard                 & $0.000\pm0.000$ & $0.000\pm0.000$ & $0.000\pm0.000$ & $0.001\pm0.000$ & $0.001\pm0.000$ & $0.001\pm0.000$ & $0.001\pm0.000$ & $0.001\pm0.000$ \\

\lasthline
\end{tabular}
\end{adjustbox}
\label{tab:fs}
\end{table*}

\section{Summary and Conclusions} 
\label{sec:summary}

We have presented a visual morphological classification of a sample of galaxies in the DEVILS/COSMOS survey with HST imaging, from the D10/ACS sample. The quality of the imaging data (HST/ACS) provides arguably the best current insight into galaxy structures and therefore the best pathway with which to explore morphological evolution. 

We summarize our results as below: 

\begin{description}

  \item[$\bullet$] By visually inspecting galaxies out to $z \sim 1.5$, we find that morphological classification becomes far more challenging at $z > 1.0$ as many galaxies ($> 40\%$) no longer adhere to the classical notion of spheroids, bulge+disk or disk systems (\citealt{Abraham01}). We see a dramatic increase in strongly disrupted systems, presumably due to interactions and gas clumps. Nevertheless, at all redshifts we find that more mass is involved in star-formation than in merging and most likely it is the high star-formation rates that are driving the irregular morphologies.
  
  \item[$\bullet$] The SMF of the D10/ACS sample in our lowest redshift bin ($z<0.25$) is consistent with previous measurements from the local Universe. 

  \item[$\bullet$] The evolution of the global SMF shows enhancement in both low- and slightly in high-mass ends. We interpret this as suggesting that at least two evolutionary pathways are in play, and that both are significantly impacting the SMF.
  
  \item[$\bullet$] Despite a slight decrease in their high-mass end, BD systems demonstrate a non-negligible growth in the low-mass end of their SMF. In the D type, we witness a significant variation in both high- and low-mass ends of the SMF. We interpret the high-mass end decrease in D systems, which is at first sight unphysical, as an indication of significant secular mass transfer through the formation of pseudo-bulges and hence an apparent mass loss as galaxies transit to the BD category.
  
  \item[$\bullet$] E galaxies experience a modest growth in their high-mass end as well as an enhancement in their low/intermediate-mass end which we interpret as a consequence of major mergers resulting in the relentless stellar mass growth of this class. 
  
  \item[$\bullet$] Despite the above shuffling of mass we find that the best regressed Schechter function parameters in the \textit{total} SMF are observed to be relatively stable from $z = 1$. This is consistent with previous studies (\citealt{Muzzin13}; \citealt{Tomczak14}; \citealt{Wright18}). Conversely, the component SMFs show significant evolution. This implication is that while stellar-mass growth is slowing, mass-transformation processes via merging and in-situ evolution are shuffling mass between types behind the scenes.
  
  \item[$\bullet$] We measured the integrated total stellar mass density and its evolution since $z = 1$ and find that approximately $32\%$ of the current stellar mass in the galaxy population was formed during the last 8 Gyr.

  \item[$\bullet$] As shown in Figure \ref{fig:MassBuildUp_LSS}, we find that the BD population dominates the SMD of the Universe within $0 \le z \le 1$ and has constantly grown by a factor of $\sim 2$ over this timeframe. On the other hand, the SMD of Ds declines slowly and eventually loses $\sim 85\%$ of it's original value until $z \sim 0.2$. On the contrary, the E population experiences constant growth of a factor of $\sim 2.5$ since $z = 1$. We observe that the extrapolation of the trends of all of our morphological SMD estimations meets GAMA measurements at $z = 0$ (see Figure \ref{fig:MassBuildUp_LSS}) except for the \textit{pure-disk} systems which is likely due to unbound distribution of their SMD (see Figure \ref{fig:Mdens_6z}). 
  
\end{description}
 
One clear outcome of our analysis is that the late Universe ($z<1$) appears to be a time of profound spheroid and bulge growth/emergence. To move forward and explore this further we conclude that to move forward it is key to decompose the double component morphological type, which significantly dominates the stellar mass density, into disks, classical bulges, and pseudo-bulges. To do this, requires robust structural decomposition which we will describe in Hashemizadeh et al. (in prep.) using our morphological classifications to guide the decomposition process.

\section{Acknowledgements}
DEVILS is an Australian project based around a spectroscopic campaign using the Anglo-Australian Telescope. The DEVILS input catalogue is generated from data taken as part of the ESO VISTA-VIDEO \citep{Jarvis13} and UltraVISTA \citep{McCracken12} surveys. DEVILS is part funded via Discovery Programs by the Australian Research Council and the participating institutions. The DEVILS website is \href{https://devilsurvey.org}{https://devilsurvey.org}. The DEVILS data is hosted and provided by AAO Data Central (\href{https://datacentral.org.au}{https://datacentral.org.au}). This work was supported by resources provided by The Pawsey Supercomputing Centre with funding from the Australian Government and the Government of Western Australia. We also gratefully acknowledge the contribution of the entire COSMOS collaboration consisting of more than 200 scientists. The HST COSMOS Treasury program was supported through NASA grant HST-GO-09822. SB and SPD acknowledge support by the Australian Research Council's funding scheme DP180103740. MS has been supported by the European Union's  Horizon 2020 research and innovation programme under the Maria Skłodowska-Curie (grant agreement No 754510), the National Science Centre of Poland (grant UMO-2016/23/N/ST9/02963) and by the Spanish Ministry of Science and Innovation through Juan de la Cierva-formacion program (reference FJC2018-038792-I). ASGR and LJMD acknowledge support from the Australian Research Council's Future Fellowship scheme (FT200100375 and FT200100055, respectively).

This work was made possible by the free and open R software (\citealt{R-Core-Team}).
A number of figures in this paper were generated using the R \texttt{magicaxis} package (\citealt{Robotham16b}). This work also makes use of the \texttt{celestial} package (\citealt{Robotham16a}) and \texttt{dftools} (\citealt{Obreschkow18}).

\section{Data Availability} 
\label{sec:AvailData}
In this work, we draw upon two datasets; the established HST imaging of the COSMOS region (\citealt{Scoville07}, \citealt{Koekemoer07}), and multiple data products produced as part of the DEVILS survey \citep{Davies18}, consisting of a spectroscopic campaign currently being conducted on the Anglo-Australian Telescope, photometry (Davies et al. in prep.), and deep redshift catalogues, and stellar mass measurements from \cite{Thorne20}.
These datasets are briefly described below.

\bibliographystyle{mnras}
\bibliography{library}

\section{Appendix A: Random samples of the morphological types.}

Here we show 49 random galaxies in each of our morphological categories (BD,D,E,H,C). We show our stamps in both HST/F814W filter and Subaru $gri$.  

\section{Appendix B: Best Schechter fit parameters.}
This table shows our best Schechter fit parameters of the total and morphological SMFs, shown in Figure \ref{fig:Mfunc_6z}.
 
\begin{table*}
\centering
\caption{Best Schechter fit parameters of \textit{total} and different morphological types in different redshift bins.}
\begin{adjustbox}{scale = 0.9}
\begin{tabular}{lcccccccc}
\firsthline \firsthline \\
$z$-bins         &  $0.0 \leq z < 0.08$  &  $0.0 \leq z < 0.25$ &  $0.25 \leq z < 0.45$ &  $0.45 \leq z < 0.60$ &  $0.60 \leq z < 0.70$ &  $0.70 \leq z < 0.80$ &  $0.80 \leq z < 0.90$ &  $0.90 \leq z \leq 1.00$ \\ \\ \hline 
                   &   \multicolumn{8}{c}{\textbf{Total (Double Schechter)}}  \\
\cline{2-9} \\
$\mathrm{log}_{10}\Phi^*_1$      & $-2.43\pm0.04$  & $-2.68\pm0.07$   & $-2.55\pm0.03$ & $-2.66\pm0.07$ & $-2.71\pm0.08$ & $-2.61\pm0.03$ & $-2.73\pm0.04$ & $-2.81\pm0.04$    \\ \\
$\mathrm{log}_{10}M^*$           & $10.65\pm0.04$  & $10.96\pm0.06$   & $10.83\pm0.03$ & $10.76\pm0.06$ & $10.79\pm0.05$ & $10.66\pm0.04$ & $10.76\pm0.04$ & $10.84\pm0.04$    \\ \\
$\alpha_1$                       & $-0.04\pm0.22$  & $-1.08\pm0.08$   & $-0.79\pm0.09$ & $-0.32\pm0.30$ & $-0.22\pm0.33$ & $0.15\pm0.20$  & $-0.02\pm0.21$ & $-0.23\pm0.20$     \\ \\
$\mathrm{log}_{10}\Phi^*_2$      & $-3.11\pm0.21$  & $-8.90\pm4.85$   & $-4.76\pm0.55$ & $-3.11\pm0.23$ & $-3.14\pm0.26$ & $-3.26\pm0.14$ & $-3.22\pm0.15$ & $-3.32\pm0.16$    \\ \\
$\alpha_2$                       & $-1.56\pm0.16$  & $-5.06\pm3.27$   & $-2.62\pm0.36$ & $-1.48\pm0.15$ & $-1.36\pm0.17$ & $-1.61\pm0.11$ & $-1.36\pm0.11$ & $-1.43\pm0.10$     \\ \\ \hline
                    &   \multicolumn{8}{c}{\textbf{Total (Single Schechter)}}  \\
\cline{2-9} \\
$\mathrm{log}_{10}\Phi^*$         & $-2.60\pm0.03$   & $-2.74\pm0.06$   & $-2.77\pm0.03$ & $-2.77\pm0.03$ & $-2.78\pm0.03$ & $-2.85\pm0.03$ & $-2.83\pm0.02$ & $-2.94\pm0.2$ \\ \\
$\mathrm{log}_{10}M^*$            & $10.93\pm0.03$   & $11.01\pm0.05$   & $11.00\pm0.03$ & $10.98\pm0.03$ & $10.99\pm0.02$ & $11.03\pm0.02$ & $11.02\pm0.02$ & $11.08\pm0.02$ \\ \\
$\alpha$                          & $-1.04\pm0.03$   & $-1.15\pm0.05$   & $-1.16\pm0.02$ & $-1.17\pm0.03$ & $-1.05\pm0.02$ & $-1.12\pm0.02$ & $-1.00\pm0.02$ & $-1.10\pm0.02$ \\ \\ \hline 
                    &   \multicolumn{8}{c}{\textbf{Double}}  \\
 \cline{2-9} \\
$\mathrm{log}_{10}\Phi^*$         & $-2.60\pm0.03$  & $-2.61\pm0.06$   & $-2.68\pm0.03$ & $-2.73\pm0.03$ & $-2.75\pm0.02$ & $-2.82\pm0.02$ & $-2.90\pm0.02$ & $-3.05\pm0.02$ \\ \\
$\mathrm{log}_{10}M^*$            & $10.75\pm0.03$  & $10.71\pm0.05$   & $10.78\pm0.03$ & $10.78\pm0.03$ & $10.77\pm0.02$ & $10.82\pm0.02$ & $10.83\pm0.02$ & $10.90\pm0.02$ \\ \\
$\alpha$                          & $-0.87\pm0.04$  & $-0.95\pm0.07$   & $-0.90\pm0.03$ & $-0.80\pm0.04$ & $-0.55\pm0.04$ & $-0.59\pm0.04$ & $-0.42\pm0.03$ & $-0.55\pm0.03$ \\ \\ \hline
                    &   \multicolumn{8}{c}{\textbf{Pure-Disk}}  \\
 \cline{2-9} \\
$\mathrm{log}_{10}\Phi^*$         & $-4.32\pm0.38$  & $-3.12\pm0.23$   & $-4.09\pm0.23$ & $-3.55\pm0.12$ & $-3.52\pm0.09$ & $-3.45\pm0.08$ & $-3.23\pm0.04$ & $-3.21\pm0.04$ \\ \\
$\mathrm{log}_{10}M^*$            & $10.82\pm0.22$  & $10.00\pm0.18$   & $10.76\pm0.14$ & $10.74\pm0.08$ & $10.78\pm0.06$ & $10.79\pm0.06$ & $10.71\pm0.03$ & $10.75\pm0.03$ \\ \\
$\alpha$                          & $-2.12\pm0.13$  & $-1.04\pm0.46$   & $-1.96\pm0.09$ & $-1.68\pm0.07$ & $-1.54\pm0.05$ & $-1.52\pm0.05$ & $-1.24\pm0.04$ & $-1.24\pm0.03$ \\ \\ \hline
                    &   \multicolumn{8}{c}{\textbf{Elliptical}}  \\
 \cline{2-9} \\
$\mathrm{log}_{10}\Phi^*$         & $-2.99\pm0.03$  & $-3.23\pm0.08$   & $-3.20\pm0.04$ & $-3.24\pm0.04$ & $-3.23\pm0.03$ & $-3.26\pm0.02$ & $-3.35\pm0.02$ & $-3.45\pm0.02$ \\ \\
$\mathrm{log}_{10}M^*$            & $10.87\pm0.04$  & $11.14\pm0.09$   & $11.00\pm0.04$ & $10.89\pm0.05$ & $10.88\pm0.04$ & $10.78\pm0.04$ & $10.87\pm0.03$ & $10.93\pm0.03$ \\ \\
$\alpha$                          & $-0.37\pm0.07$  & $-0.71\pm0.10$   & $-0.61\pm0.05$ & $-0.40\pm0.08$ & $-0.25\pm0.07$ & $0.09\pm0.09$  & $-0.07\pm0.07$ & $-0.03\pm0.07$ \\ \\ \hline
                    &   \multicolumn{8}{c}{\textbf{Compact}}  \\
 \cline{2-9} \\
$\mathrm{log}_{10}\Phi^*$         & $-$  & $-19.91\pm45.11$   & $-4.01\pm1.14$ & $-9.70\pm62.46$ & $-4.76\pm0.89$ & $-3.31\pm0.20$ & $-4.72\pm0.32$ & $-4.69\pm0.24$ \\ \\
$\mathrm{log}_{10}M^*$            & $-$  & $8.37\pm0.74$      & $9.79\pm0.51$  & $12.71\pm32.57$ & $10.37\pm0.41$ & $9.67\pm0.17$  & $10.73\pm0.20$ & $10.86\pm0.16$ \\ \\
$\alpha$                          & $-$  & $17.85\pm35.50$    & $-2.09\pm1.37$ & $-2.90\pm0.28$  & $-2.29\pm0.42$ & $-1.16\pm0.62$ & $-1.79\pm0.15$ & $-1.75\pm0.11$ \\ \\ \hline
                    &   \multicolumn{8}{c}{\textbf{Hard}}  \\
 \cline{2-9} \\
$\mathrm{log}_{10}\Phi^*$         & $-5.34\pm0.94$  & $-5.42\pm2.50$   & $-6.72\pm1.60$ & $-5.06\pm0.40$ & $-5.36\pm0.37$ & $-4.73\pm0.24$ & $-4.42\pm0.12$ & $-4.29\pm0.10$ \\ \\
$\mathrm{log}_{10}M^*$            & $11.06\pm0.70$  & $11.00\pm1.56$   & $12.34\pm1.43$ & $11.40\pm0.33$ & $11.74\pm0.33$ & $11.22\pm0.19$ & $11.21\pm0.11$ & $11.16\pm0.09$ \\ \\
$\alpha$                          & $-1.35\pm0.69$  & $-2.01\pm0.75$.  & $-1.97\pm0.14$ & $-1.67\pm0.13$ & $-1.64\pm0.10$ & $-1.55\pm0.10$ & $-1.27\pm0.07$ & $-1.29\pm0.06$ \\ \\

\lasthline
\end{tabular}
\end{adjustbox}
\label{tab:Morph_MF_par}
\end{table*}

\section{The non-LSS-corrected evolution of the SMD} 
\label{sec:SMD_evol_noLSS}

\begin{figure*}
\centering
  \centering
  \includegraphics[width=\textwidth]{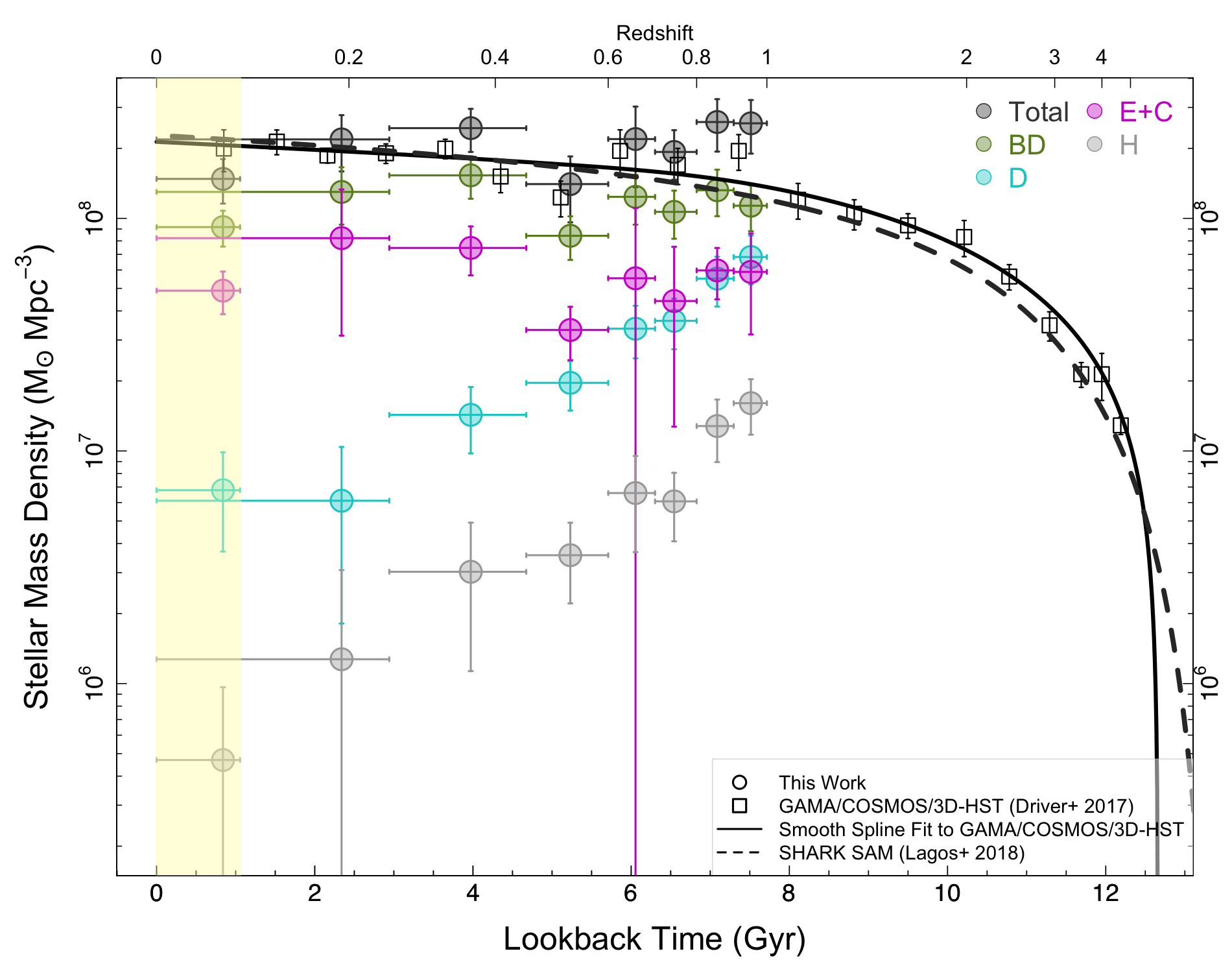}
  \caption{The evolution of the stellar mass density (SMD) of total and morphological types before applying the LSS corrections. Highlighted region shows the epoch covered by the GAMA data.}
  \label{fig:SMD_evol_noLSS}
\end{figure*}

Figure \ref{fig:SMD_evol_noLSS} shows the evolution of the integrated stellar mass density, $\rho_*$ before we apply our large scale structure corrections (reported in Table \ref{tab:rho} and shown in Figure \ref{fig:LSS_cor}). This is to further confirm that the corrections do not derive the overall trends that we find in Figure \ref{fig:MassBuildUp_LSS} as explained in Section \ref{sec:rho}.

\label{lastpage}
\end{document}